\documentclass[preprint,aps,pra]{revtex4}
\usepackage{tabularx}
\usepackage{dcolumn}
\usepackage{graphics}
\usepackage{bm}
\usepackage[dvips]{graphicx}
\usepackage{tabularx}
\usepackage{amsmath}
\usepackage{amssymb}
\usepackage{longtable}
\usepackage{verbatim}
\usepackage{float} 
\usepackage{fancyhdr} 
\usepackage{color}

\begin{document}
\pagestyle{fancy}
\rhead{\thepage}
\chead{}
\lhead{}
\cfoot{}
\rfoot{}
\lfoot{}


\title{Decays and spectrum of bottom and bottom strange mesons}
\author{Ishrat Asghar}\affiliation{Centre For High Energy Physics, Punjab
University, Lahore(54590), Pakistan. }
\author{Bilal Masud}\affiliation{Centre For High Energy Physics, Punjab
University, Lahore(54590), Pakistan. }
\author{E. S. Swanson}\affiliation{University of Pittsburgh, Pittsburgh PA 15260, USA.}
\author{Faisal Akram}\affiliation{Centre For High Energy Physics, Punjab University,
Lahore(54590), Pakistan.} 
\author{M. Atif Sultan}\affiliation{Centre For High Energy Physics, Punjab University,
Lahore(54590), Pakistan.}

\begin{abstract}
The strong decay amplitudes and radiative partial widths of orbital and radially excited states of $B$ and $B_s$ mesons are presented. These results are obtained with a nonrelativistic potential quark model, the nonrelativistic reduction of the electromagnetic transition operator, and the ``$^3P_0$" model of strong decays. The predictions are compared to experiment where possible and assignments for the recently discovered states,
$B_1(5721)$, $B_2^*(5747)$, $B_J(5840)$, $B_J(5970)$, $B_{s1}(5830)$, and $B_{s2}^*(5840)$, are made.
\end{abstract}
\maketitle

\section{Introduction}

In 2007 the DO collaborations\cite{DO 2007} announced the discoveries of bottom mesons $B_1(5721)$ and $B_2^*(5747)$. This was followed with sightings of the $B_{s1}(5830)$ and $B_{s2}^*(5840)$ states at LHCb\cite{LHCb 2013} in 2013. More recently CDF\cite{CDF 2014} and LHCb\cite{lhcb2015} have found evidence for excited bottom mesons, $B_J(5970)$ and $B_J(5940)$.
These recent discoveries of bottom and bottom-strange mesons motivate model computations in these sectors. The immediate goal is to assist in identifying these states, while a longer term goal is to determine the limits of potential models as useful descriptions of low lying hadrons. To this end we construct a reasonably ``standard" nonrelativistic potential quark model and fit its parameters to the known $B$ and $B_s$ mesons.  The philosophy underpinning this type of modelling is the adiabatic separation of ``slow" constituent quark degrees of freedom from ``fast" gluonic degrees of freedom~\cite{nosheen11,atif 2014,morningstar,Braaten}. The spectrum and wavefunctions that emerge from this computation are used to compute E1 and M1 radiative transitions. Finally, the ``$^3P_0$" model of strong decays, wherein it is assumed that quark-antiquark pairs are created with vacuum quantum numbers, is used to compute open flavour decays of a collection of  low lying $B$ and $B_s$ states.

The organization of the paper is as follows. First, we
describe the potential model used to calculate the mass spectrum of the bottom and bottom-strange mesons. Mixed states are discussed in Sec. \ref{mixed states} along with conventions used in the literature.
In Sec. \ref{sect:decay}, we review the $^3P_0$ decay model evaluation of strong decay amplitudes using simple harmonic oscillator (SHO) wave functions.
E1 and M1 radiative transitions are calculated in Sec. \ref{radiative transitions}.
The results of these computations are compared with available experimental data in Sec. \ref{results}, while our conclusions are given in Sec. \ref{conclusions}.

\section{Constituent Quark Model for $B$ and  $B_s$ States}
\label{spectrum}

The chief assumptions underpinning our model are nonrelativistic kinematics, a ``Coulomb+linear" central potential, and spin-dependent corrections generated from vector gluon exchange and an effective scalar confinement interaction. While it is relatively simple to incorporate relativistic kinematics for the quarks, these effects can be largely subsumed into the constituent quark mass, and in fact, there is evidence that this is a preferred approach\cite{Lucha 1992, Gunion 1975}.


The potential used in this paper is given by

\begin{eqnarray}\label{potential}
V_{q\overline{q}}(r) &=& -\frac{4\alpha_s}{3r} + br + \frac{32\pi \alpha_s}{9m_qm_{\overline{q}}}(\frac{\sigma}{\sqrt{\pi}})^3 e^{-\sigma^2r^2}\textbf{S}_q\cdot \textbf{S}_{\overline{q}}+ \frac{4\alpha_s}{m_q m_{\overline{q}}r^3} T \nonumber \\
 && + \left( \frac{\textbf{S}_q}{8 m_q^2} + \frac{\textbf{S}_{\overline{q}}}{8 m_{\overline{q}}^2}\right) \cdot \textbf{L} \left(\frac{4 \alpha_s}{3 r^3} - \frac{b}{r}\right) + \frac{\textbf{S}_q + \textbf{S}_{\overline{q}}}{4 m_q m_{\overline{q}}} \cdot \textbf{L} \, \frac{4 \alpha_s}{3 r^3}.
\end{eqnarray}
Here $\alpha_s$ is the strong coupling constant, $b$ is the string tension,
and $T$ is the tensor operator

\begin{equation}
T = \textbf{S}_q\cdot \hat r \, \textbf{S}_{\overline{q}}\cdot \hat r - \frac{1}{3} \textbf{S}_q\cdot \textbf{S}_{\overline{q}}
\end{equation}
with diagonal matrix elements given by
\begin{equation}
 T= \left\{
      \begin{array}{ll}
        -\frac{L}{6(2L+3)}&\hspace{0.4cm} J=L+1 \\
       \frac{1}{6}&\hspace{0.4cm} J=L \\
        -\frac{(L+1)}{6(2L-1)}&\hspace{0.4cm} J=L-1.
      \end{array}
    \right.
\end{equation}

The parameters used in this potential for $B$ and $B_s$ mesons are taken to be $\sigma=0.84\;GeV$, $\alpha_s(B)=0.775$, $\alpha_s(B_s)=0.642$ and $b=0.0945$ GeV$^2$. These values were obtained by fitting the masses of eight experimentally known states of $B$ and $B_s$ mesons (these are listed in $4^{th}$ and $7^{th}$ columns of Table \ref{B table}). The light quark masses, $m_{u/d}$ and $m_s$,  were obtained from fits to light mesons. Finally, the bottom quark mass of Ref. \cite{nosheen 2015} was used. These masses are
\begin{eqnarray}
  m_u=m_d=0.325\;\textrm{GeV},  \nonumber \\
  m_s=0.422\;\textrm{GeV}, \nonumber \\
  m_b=4.825\;\textrm{GeV}.
\end{eqnarray}

The radial Schr\"{o}dinger equation was solved with the shooting method\cite{atif 2014};  results for the $B$ and $B_s$ spectrum are reported in Table \ref{B table}.

\begin{table}[H]
\centering
\renewcommand{\arraystretch}{0.6}
\caption{Masses of ground and radially excited states of $B$ and $B_s$ mesons.}
\begin{tabular}{c c c c c c c}
  \hline\hline
  n  & \hspace{0.3cm} Meson  & \hspace{0.3cm} Our calculated & \hspace{0.3cm} Expt. Mass \cite{PDG2014}& \hspace{0.3cm} Meson & \hspace{0.3cm} Our calculated & \hspace{0.3cm} Expt. Mass \cite{PDG2014}\\
   &  &Mass (GeV) & (GeV) & & Mass (GeV) & (GeV)\\
   \hline
1S & $B(1^1S_0)$  & 5.2675  & $5.27926\pm 0.00017$ & $B_s(1^1S_0)$ & 5.37697 & $5.36677\pm 0.00024$ \\
   & $B(1^3S_1)$  & 5.32949 & $5.3252\pm 0.0004$   & $B_s(1^3S_1)$ & 5.42194 & $5.4154^{+0.0024}_{-0.0021}$ \\
\hline
2S & $B(2^1S_0)$  & 5.87731 &...                   & $B_s(2^1S_0)$ & 5.92868 &... \\
   & $B(2^3S_1)$  & 5.90536 & ...                  & $B_s(2^3S_1)$ & 5.94895 & ...\\
\hline
3S & $B(3^1S_0)$  & 6.28798 & ...                  & $B_s(3^1S_0)$ & 6.30477 & ...\\
   & $B(3^3S_1)$  & 6.30821 & ...                  & $B_s(3^3S_1)$ & 6.31925 & ...\\
\hline
4S & $B(4^1S_0)$  & 6.63061 &...                   & $B_s(4^1S_0)$ & 6.61943 &... \\
   & $B(4^3S_1)$  & 6.64701 & ...                  & $B_s(4^3S_1)$ & 6.63109 & ...\\
\hline
1P & $B(1^3P_0)$  & 5.70364 & ...                  & $B_s(1^3P_0)$ & 5.76968 & ...\\
   & $B(1^3P_2)$  & 5.76895 & $5.743\pm 0.005$     & $B_s(1^3P_2)$ & 5.82153 & $5.83996\pm 0.0002$ \\
   & $B(1\;P_1)$  & 5.75512 & $5.7235\pm 0.002$    & $B_s(1\;P_1)$ & 5.80299 & $5.8287\pm 0.0004$\\
   & $B(1\;P'_1)$ & 5.73920 & ...                  & $B_s(1\;P'_1)$& 5.80095 & ... \\
   & $\phi_{1P}$  & $35.3^\circ$ &                 &  $\phi_{1P}$  & $35.3^\circ$  \\
\hline
2P & $B(2^3P_0)$  & 6.12867 & ...                  & $B_s(2^3P_0)$ & 6.15958 & ...\\
   & $B(2^3P_2)$  & 6.18965 &...                   & $B_s(2^3P_2)$ & 6.20834 &... \\
   & $B(2\;P_1)$  & 6.17525 &...                   & $B_s(2\;P_1)$ & 6.19639 & ...\\
   & $B(2\;P'_1)$ & 6.1614  & ...                  & $B_s(2\;P'_1)$& 6.18615 & ... \\
   & $\phi_{2P}$  & $35.3^\circ$ &                 &  $\phi_{2P}$  & $35.3^\circ$  \\
\hline
3P & $B(3^3P_0)$  & 6.48027 & ...                  & $B_s(3^3P_0)$ & 6.48284 & ...\\
   & $B(3^3P_2)$  & 6.53949 &...                   & $B_s(3^3P_2)$ & 6.53036 &... \\
   & $B(3\;P_1)$  & 6.52499 &...                   & $B_s(3\;P_1)$ & 6.51843 & ...\\
   & $B(3\;P'_1)$ & 6.51159 & ...                  & $B_s(3\;P'_1)$& 6.50828 & ... \\
   & $\phi_{3P}$  & $35.3^\circ$ &                 &  $\phi_{3P}$  & $35.3^\circ$  \\
\hline
1D & $B(1^3D_1)$  & 6.02239 & ...                  & $B_s(1^3D_1)$ & 6.05682 & ...\\
   & $B(1^3D_3)$  & 6.03076 &...                   & $B_s(1^3D_3)$ & 6.06293 &... \\
   & $B(1\;D_2)$  & 6.03122 &...                   & $B_s(1\;D_2)$ & 6.06403 & ...\\
   & $B(1\;D'_2)$ & 6.02649 & ...                  & $B_s(1\;D'_2)$& 6.05935 & ... \\
   & $\phi_{1D}$  & $39.2^\circ$ &                 &  $\phi_{1D}$  & $39.2^\circ$  \\
\hline
2D & $B(2^3D_1)$  & 6.38438 & ...                  & $B_s(2^3D_1)$ & 6.39012 & ...\\
   & $B(2^3D_3)$  & 6.39544 &...                   & $B_s(2^3D_3)$ & 6.39853 &... \\
   & $B(2\;D_2)$  & 6.39323 &...                   & $B_s(2\;D_2)$ & 6.39741 & ...\\
   & $B(2\;D'_2)$ & 6.39075 & ...                  & $B_s(2\;D'_2)$& 6.39453 & ... \\
   & $\phi_{2D}$  & $39.2^\circ$ &                 &  $\phi_{2D}$  & $39.2^\circ$  \\
\hline
1F & $B(1^3F_2)$  & 6.25908 & ...                  & $B_s(1^3F_2)$ & 6.27319 & ...\\
   & $B(1^3F_4)$  & 6.252   &...                   & $B_s(1^3F_4)$ & 6.26679 &... \\
   & $B(1\;F_3)$  & 6.2637  &...                   & $B_s(1\;F_3)$ & 6.27694 & ...\\
   & $B(1\;F'_3)$ & 6.24939 & ...                  & $B_s(1\;F'_3)$& 6.26467 & ... \\
   & $\phi_{1F}$  & $40.9^\circ$ &                 &  $\phi_{1F}$  & $40.9^\circ$  \\
  \hline\hline
\end{tabular}
\label{B table}
\end{table}

\subsection{Mixed States}\label{mixed states}

Heavy-light mesons are not charge conjugation eigenstates and so mixing can occur between states with $J=L$ and $S=0,1$. For example, the two $J^P=1^+$ states are a mixture of $n{}^1P_1$ and $n{}^3P_1$ states. Thus
\begin{eqnarray}
  |B(1\;P_1)\rangle &=& +\cos(\phi_{1P})|1\;^1P_1\rangle+\sin(\phi_{1P})|1\;^3P_1\rangle \nonumber \\
  |B(1\;P_1')\rangle &=&-\sin(\phi_{1P})|1\;^1P_1\rangle+\cos(\phi_{1P})|1\;^3P_1\rangle \nonumber
\end{eqnarray}
where $\phi_{1P}$ is the mixing angle.
For D-waves mixing occurs between $^3D_2$ and $^1D_2$ states, with a mixing angle denoted as $\phi_{nD}$ for principal quantum number $n$. In the heavy quark limit $m_Q\rightarrow \infty$ the mixing angles become\cite{ebert 2010}
\begin{equation}\label{mixing angle}
  \phi_{m_Q\rightarrow \infty}=\arctan(\sqrt{\frac{L}{L+1}}).
\end{equation}
This implies $\phi_{1P}=\phi_{2P}=35.3^{\circ}$ and $\phi_{1D}=\phi_{2D}=39.2^{\circ}$. We have confirmed that these angles are close to those produced by our potential model and hence use them in the subsequent work.

\section{Strong Decay Model}
\label{sect:decay}

Strong decays will be computed using the $^3P_0$ model for quark-antiquark pair production.
This model dates back to Micu in 1969 \cite{micu 1969}; it assumes that quark and antiquarks are produced with vacuum quantum numbers with a strength that is determined by experiment. The model was developed extensively by
  Le Yaouanc and collaborators\cite{yaounac 1973,yaounac 1974,yaounac 1975} and applied successfully to many meson decays. The $^3P_0$ model has also been used to describe baryon decays\cite{roberts 1992,roberts 1998,yaounac 1974,capstick 1992,capstick 1994} and for diquonia decays to baryon plus antibaryon \cite{roberts 1990}. Older studies of strong decays considered an alternative phenomenological model in which quark-antiquark pairs are produced with $^3S_1$ quantum numbers \cite{alcock 1984,kumano 1988}; however, this possibility is disfavoured by measured ratios of partial wave amplitudes~\cite{geiger 1994}.

 A comprehensive study of light meson strong decays in the $^3P_0$ model was made by Barnes \textit{et al.} in Ref. \cite{barnes 1997}. Simple harmonic oscillator (SHO) meson wave functions were used in this work. The same assumptions were used  for strong decays in the  $n\overline{s}$, $s\overline{s}$, $c\overline{c}$, $c\overline{n}$, $b\overline{n}$, $b\overline{s}$ $(n=u$ or $d)$ sectors~\cite{barnes 2003,barnes 2005,close 2005,godfrey 2016}.  We adopt the same approach here, but fit the SHO wavefunction parameter (denoted $\beta$ in the following) to the wavefunctions obtained above.

\begin{figure}[!h]
\hspace{5cm}\includegraphics[width=14cm]{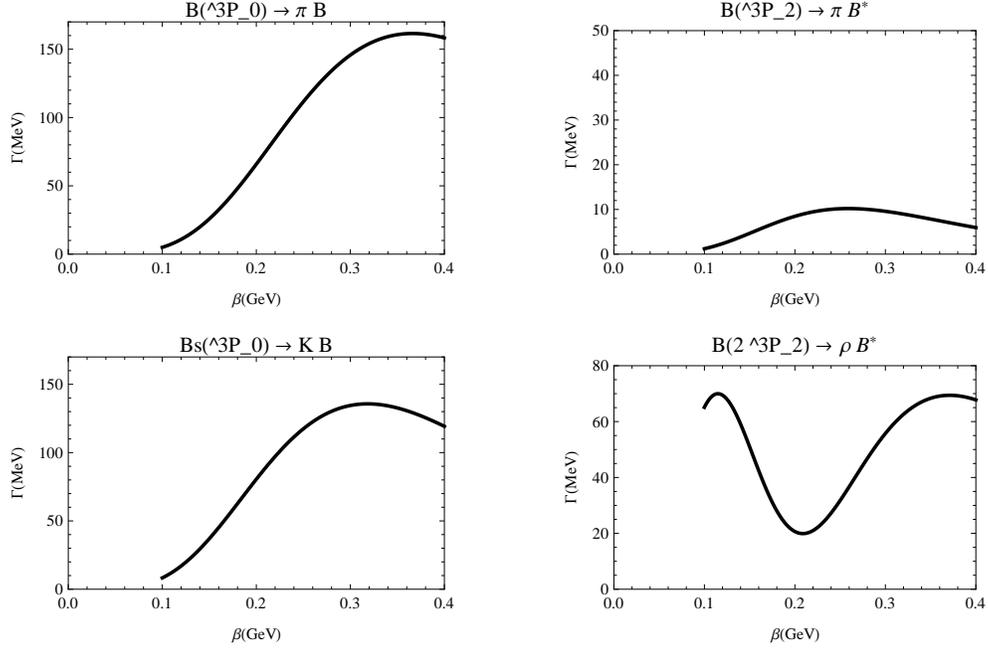}
\caption{Partial widths of some of the strong decays of $B$ and $B_s$ mesons in the $^3P_0$ model are plotted as a function of $\beta$ for the initial meson.}
\label{betadependence}
\end{figure}

The dependence of typical strong decays on the choice of the SHO parameter $\beta$ for the $B$ meson is illustrated in Fig. \ref{betadependence}, which indicates sensitivities of $d\Gamma/d\beta \approx 0.5$ or less.

\begin{figure}[!h]
\centering
\includegraphics[width=15cm]{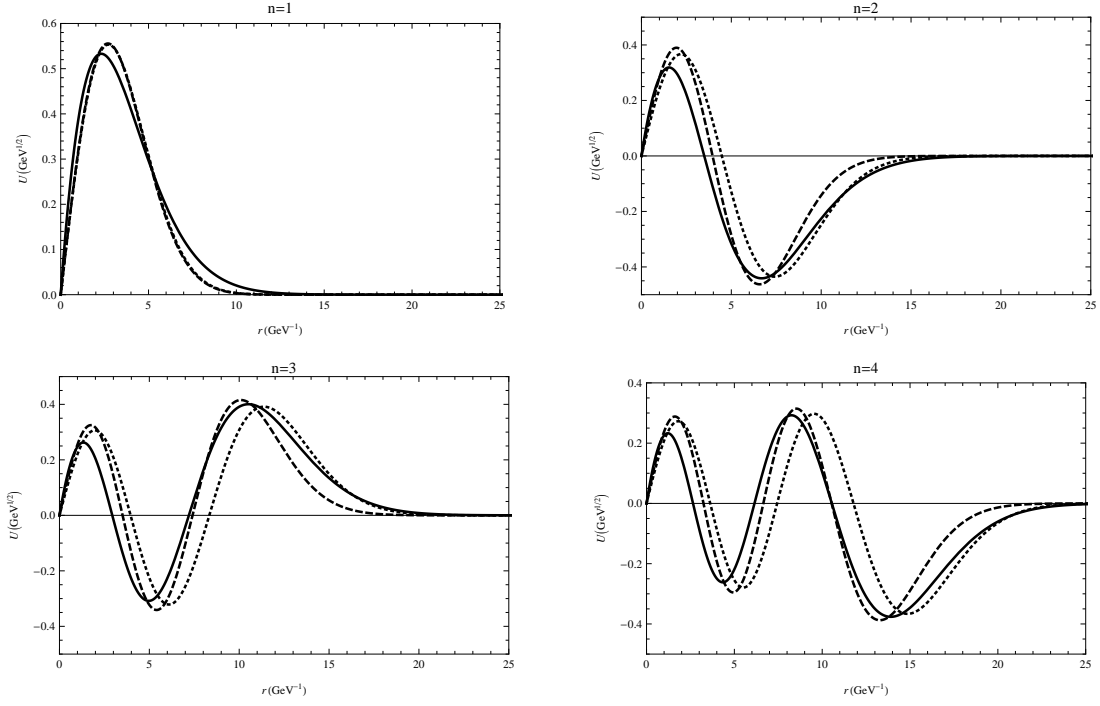}
\caption{Radial Wavefunctions of $n^3S_1$ states of $B$ meson with $n=1,2,3,4$. Solid: full model, dotted: SHO with RMS $\beta$, dashed: SHO with fit $\beta$.}
\label{beta graphs}
\end{figure}

In Refs.\cite{barnes 2005,godfrey 2016} the $\beta$ parameter was obtained by equating the root mean square (RMS) radius of a harmonic oscillator wavefunction to the RMS radius of the analogous quark model wavefunction. In this work we choose to
 obtain  $\beta$ by fitting the SHO wavefunction to the quark model wavefunction. Typical results are shown in  Fig.(\ref{beta graphs}), which compares wavefunctions obtained with all three methods. SHO wavefunction parameters and the error,  $\int(U-U_{SHO})^2\;dr$, for these cases are reported in  Table \ref{error table}. Both methods assume that a typical momentum dominates the decay amplitude with one optimises to the state RMS radius and the other to a global fit to the wavefunction.

\begin{table}[h!]
\centering
\renewcommand{\arraystretch}{0.7}
\caption{Quark model and SHO wavefunction errors, $\int(U - U_{SHO})^2\;dr$, for the RMS and fit methods.}
\begin{tabular}{c c c c c }
\hline
\hline
Meson\hspace{.1 in}&\hspace{.1 in} Quantity \hspace{.1 in}& By equating rms radius of the     &  \hspace{.3 in} By fitting $\beta$  in the SHO form \\
                &               & \hspace{.1 in} realistic and SHO wavefunctions & \hspace{.3 in} to the realistic wavefunction\\
\hline
$B(1 ^3S_1)$    &$\beta(GeV)$   &  0.368   & 0.371    \\
                &Error          &  0.0138  & 0.0137    \\
\hline
$B(2 ^3S_1)$    &$\beta(GeV)$   &  0.274   & 0.309    \\
                &Error          &  0.146   & 0.069    \\
\hline
$B(3 ^3S_1)$    &$\beta(GeV)$   &  0.242   & 0.273    \\
                &Error          &  0.267   & 0.090    \\
\hline
$B(4 ^3S_1)$    &$\beta(GeV)$   &  0.225   & 0.252    \\
                &Error          &  0.398   & 0.113    \\
\hline
\hline
\end{tabular}
\label{error table}
\end{table}

With the wavefunctions specified, we now move on to the decay model definition.
The interaction Hamiltonian for the $^3P_0$ model is obtained from the nonrelativistic limit of
\begin{equation}\label{hamiltonian}
H_I=2 m_q \gamma\int d^3\textbf{x}\;\overline{\psi}_q(\textbf{x}) \psi_q(\textbf{x}),
\end{equation}
where $\gamma$ is a coupling to be determined.
The connection with quark-antiquark production can be seen through the familiar second quantized form of the Dirac quark field $\psi$
\begin{equation}
  \psi_q(\textbf{x})=\int\frac{d^3\textbf{k}}{(2\pi)^3}[u(\textbf{k},s)b(\textbf{k})+v(-\textbf{k},s)d^{\dag}(-\textbf{k})]
  e^{i\textbf{k}.\textbf{x}}.\label{psi}
\end{equation}
In the notation of Eq.(\ref{psi}), quark-antiquark production in an open flavour strong meson decay $A\rightarrow B+C$ is described by the $b^{\dag}d^{\dag}$ term in $H_I$ of Eq.(\ref{hamiltonian}). The pair production strength $\gamma$ is a dimensionless constant and is determined to be approximately $\gamma = 0.35$ in a fit to the known $c\overline{c}$ states above open-charm threshold\cite{barnes 2005}.

In this paper, we use a modified version of the  pair creation strength that replaces $\gamma$ with
\begin{equation}
  \gamma^{\textmd{eff}}=\frac{m_{u/d}}{m}\;\gamma, \nonumber
\end{equation}
where $m$ is the mass of the produced quark \cite{kalasnikova,Ferretti1,Ferretti2}.
This mechanism suppresses those diagrams in which a heavy $q\overline{q}$ pair is created\cite{Ferretti2} and is equivalent to fixing the prefactor in Eq. \ref{hamiltonian} to be $2 m_{u/d}$.

To calculate the decay rate of process $A\rightarrow B+C$, we evaluate the matrix element $\langle BC|H_I|A\rangle$ by using quark model $q\overline{q}$ states for the initial and final mesons of the form
\begin{eqnarray}\label{state function}
  |A\rangle = |\mathbf{A};nJM[LS];II_z\rangle=  \int\;d^3\mathbf{a}d^3\overline{\mathbf{a}}\;
  \delta(\mathbf{A}-\mathbf{a}-\overline{\mathbf{a}})\;
  \phi_{nL}(\frac{m_{\overline{a}}\mathbf{a}-m_a\overline{\mathbf{a}}}{m_{a}+m_{\overline{a}}})\; \\ \nonumber
  X^{JM[LS]}_{c,s,f;\overline{c},\overline{s},\overline{f}}\;
  Y_{LM_L}(\hat{k})\;
  b^{\dagger}_{c,s,f}(\mathbf{a})\;
  d^{\dagger}_{\overline{c},\overline{s},\overline{f}}(\overline{\mathbf{a}})\;|0\rangle,
\end{eqnarray}
where $X^{JM[LS]}_{c,s,f;\overline{c},\overline{s},\overline{f}}$ is a matrix given by
\begin{equation}
  X^{JM[LS]}_{c,s,f;\overline{c},\overline{s},\overline{f}}=\frac{\delta_{c\overline{c}}}{\sqrt{3}}\;
  \Xi_{f,\overline{f}}^{I,I_z}\;\langle\frac{1}{2}s,\frac{1}{2}\overline{s}|SM_S\rangle\;\langle SM_S,LM_L|JM\rangle .\nonumber
\end{equation}
A sum over repeated indices is understood in Eq.(\ref{state function}). In the last equation
$\Xi_{f,\overline{f}}^{I,I_z}$ is a flavor wavefunction and $\phi$ is the spatial wavefunction that depends on the momenta $\mathbf{a}$ and $\overline{\mathbf{a}}$ of the quark and antiquark with masses $m_{a}$ and $m_{\overline{a}}$.

\begin{figure}[!h]
\hspace{5cm}\includegraphics[width=12cm]{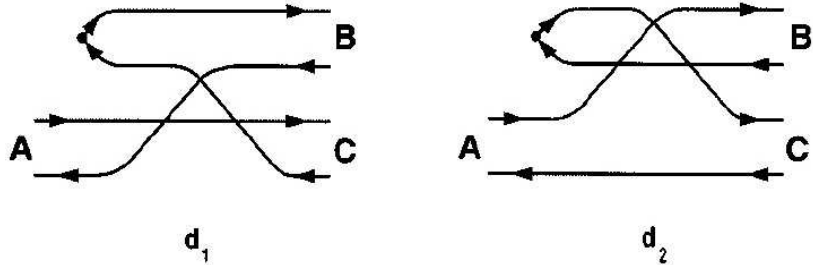}
\caption{Decay diagrams in the $^3P_0$ Model.}
\label{3p0diagram}
\end{figure}

In general four diagrams contribute to each decay amplitude. Two of these correspond to OZI suppressed decays and are disallowed by momentum conservation. The remaining diagrams are shown in Fig.(\ref{3p0diagram}).
As in Ref.~\cite{ackleh 1996}, we call the resulting diagram $d_1$ if the produced quark goes into meson $B$ and $d_2$ if it goes into meson $C$.

The matrix element for each diagram factorises according to
\begin{equation} \label{factors}
  \langle BC|H_I|A\rangle=I_{signature}I_{flavour}I_{spin}I_{space}.
\end{equation}
The first factor is, in the notation of Ref.~\cite{ackleh 1996},
\begin{equation}\label{operator factor}
 I_{signature}= \langle 0|b_\textbf{c} d_{\overline{\textbf{c}}}b_\textbf{b} d_{\overline{\textbf{b}}}b^{\dag}_\textbf{k} d^{\dag}_{-\textbf{k}}d^{\dag}_{\overline{\textbf{a}}}b^{\dag}_\textbf{a}|0\rangle,
\end{equation}
which evaluates to
\begin{equation}
  I_{signature}=[\delta(\overline{\textbf{c}}+\textbf{k})\delta(\overline{\textbf{b}}-\overline{\textbf{a}})-\delta(\overline{\textbf{b}}+\textbf{k})\delta(\overline{\textbf{c}}-\overline{\textbf{a}})]
[\delta(\textbf{b}-\textbf{k})\delta(\textbf{c}-\textbf{a})-\delta(\textbf{c}-\textbf{k})\delta(\textbf{b}-\textbf{a})].
\end{equation}

The decay vertex is unity in flavour space, hence the flavour factor $I_{flavour}$ is simply an isospin overlap matrix element. For a decay process having several charge channels (for example $B^{+}\rightarrow \pi^+ B^0$ and $B^{+}\rightarrow \pi^0 B^+$) the sum over all the charge channels is performed. This is summarised as a multiplicative flavour factor $\mathcal{F}$. The flavour factor $I_{flavour}(d_1)$ for each process discussed in this paper is zero. For all the processes discussed in this work, $I_{flavor}(d_2)$ and $\mathcal{F}$ are reported in Table \ref{flavor factor}.
\begin{table}[h!]
\centering
\begin{tabular}{c c c c c }
\hline
\hline
Generic Decay&Subprocess&$I_{flavor}(d_2)$&$\mathcal{F}$\\
\hline
$B\rightarrow \pi B $     &  $B^{+}\rightarrow \pi^+ B^0$    & 1            & 3/2  \\
$B\rightarrow \eta B $    &  $B^{+}\rightarrow \eta B^+$     & 1/2          & 1  \\
$B\rightarrow \omega B $  &  $B^{+}\rightarrow \omega B^+$   & $1/\sqrt{2}$ & 1  \\
$B\rightarrow \rho B $    &  $B^{+}\rightarrow \rho^+ B^0$   & 1            & 3/2  \\
$B\rightarrow K B_s $     &  $B^{+}\rightarrow K^+ B_s^0$    & 1            & 1  \\
$B_s\rightarrow K B $     &  $B_s^{0}\rightarrow K^- B^+$    & 1            & 2  \\
$B_s\rightarrow \eta B_s$ &  $B_s^{0}\rightarrow \eta B_s^0$ & $1/\sqrt{2}$ & 1  \\
\hline\hline
\end{tabular}
\caption{Flavour Factors}\label{flavor factor}
\end{table}

Substituting Eqs.(\ref{hamiltonian}-\ref{state function}) in $\langle BC|H_I|A\rangle$, we get the (remaining) spin and space factor of the diagram $(d_2)$ for the general process $q_1\overline{b}\rightarrow q_1\overline{q}_2+q_2\overline{b}$, where $q_1\;or\;q_2=(u,d,s)$, as
\begin{eqnarray}\label{space factor 1}
  I_{spin+space}(d_2)=-2m_q\gamma\int d^3\textbf{a}\;d^3\overline{\textbf{a}}\;d^3\textbf{b}\;d^3\overline{\textbf{b}}\;d^3\textbf{c}\;d^3\overline{\textbf{c}}\;
  \phi_A(\frac{m_b\textbf{a}-m_{q_1}\overline{\textbf{a}}}{m_b+m_{q_1}})\;\delta(\textbf{A}-\textbf{a}-\overline{\textbf{a}}) \nonumber\\
\cdot \phi_B^*(\frac{m_{q_2}\textbf{b}-m_{q_1}\overline{\textbf{b}}}{m_{q_1}+m_{q_2}})\;\delta(\textbf{B}-\textbf{b}-\overline{\textbf{b}})\;
  \phi_C^*(\frac{m_b\textbf{c}-m_{q_2}\overline{\textbf{c}}}{m_b+m_{q_2}})\;\delta(\textbf{C}-\textbf{c}-\overline{\textbf{c}}) \nonumber \\
  \cdot \int\frac{d^3\textbf{k}}{(2\pi)^3}\;\delta(\overline{\textbf{a}}-\overline{\textbf{c}})\;\delta(\textbf{a}-\textbf{b})\;
  \delta(\textbf{k}-\textbf{c})\;\delta(-\textbf{k}-\overline{\textbf{b}})\;\langle s|\overline{u}(\textbf{k},s)v(-\textbf{k},s)|\overline{s}\rangle. \hspace{0.3cm}
\end{eqnarray}
The matrix element
$\langle s|\overline{u}(\textbf{k},s)v(-\textbf{k},s)|\overline{s}\rangle$ is $\frac{1}{m_q}\langle s|\overrightarrow{\sigma}|\overline{s}\rangle\cdot\textbf{k}$ in the nonrelativistic limit. Here $\langle s|\overrightarrow{\sigma}|\overline{s}\rangle$ is a spin matrix element which can be calculated by using Eqs. B14-B16 of Ref. \cite{ackleh 1996}.


Performing the integrations gives three expressions of the same form in terms of the spherical components of $\textbf{B}$. Rather than calculating each, we compute one of them and use the Wigner-Eckart theorem to find the others.

Setting $\vec k = k_z$, $\textbf{A}=0$, and $\textbf{C}=-\textbf{B}$ gives
\begin{eqnarray} \label{space factor}
  I_{space}(d_2)=-2\gamma\int \frac{d^3\textbf{k}}{(2\pi)^3}\;k_z\;
  \phi_A(\textbf{B}+\textbf{k})\;
  \phi_B^*(\frac{m_{q_2}}{m_{q_1}+m_{q_2}}\textbf{B}+\textbf{k})\;
  \phi_C^*(\textbf{k}+\frac{m_{q_2}}{m_b+m_{q_2}}\textbf{B}).
\end{eqnarray}

The three-dimensional SHO wavefunctions appearing in this expression have the form\cite{barnes 1997,ackleh 1996}
\begin{equation}\label{SHO wave function}
  \phi_{nlm}(\textbf{k})=(2\pi)^{3/2}(-i)^{2n+l}N_{nlm}k^lY_{lm}(\hat{k})L_n^{l+1/2}(\beta^{-2}k^2)e^{-\frac{1}{2}\beta^{-2}k^2},
\end{equation}
where $\beta$ is the SHO parameter, $N_{nlm}$ is a normalization constant, and $L_n^{l+1/2}$ is the associated Legendre polynomial.

Finally, the $^3P_0$ amplitude can be combined with relativistic phase space to express the decay width for ($A\rightarrow B+C$) in the form\cite{ackleh 1996}
\begin{equation}\label{gammaTOT}
\Gamma_{A\rightarrow BC}=2\pi\frac{P E_BE_C}{M_A}\sum_{LS} |\mathcal{M}_{LS}|^2,
\end{equation}
where $P=|\textbf{B}|=|\textbf{C}|$, $M_A$ is the mass of the initial meson, and $E_B$ and $E_C$ are the energies of the final mesons $B$ and $C$ respectively. Where available, we use experimental masses\cite{PDG2016}; otherwise we employ the theoretical masses of Table \ref{B table}.

The partial wave amplitudes referred to in Eq. \ref{gammaTOT} are given by
\begin{equation}
\mathcal{M}_{LS} = \sum_{M_L M_S M_B M_C} \langle L M_L \, S M_S| J_A M_A\rangle \, \langle J_B M_B\, J_C M_C | S M_S\rangle \, \int Y^*_{L M_L}(\hat B)\, \mathcal{M}(\textbf{B})\, d\Omega_B
\end{equation}
and $\mathcal{M}$ is the decay amplitude of Eq. \ref{factors}.

\section{Radiative Transitions of $B$ and $B_s$ Mesons}\label{radiative transitions}
\subsection{E1 Radiative Transitions}

E1 radiative transition partial widths are computed  with the following expression
\begin{equation}
  \Gamma(n^{2S+1}L_J\rightarrow n'^{2S'+1}L'_{J'}+\gamma)=\frac{4}{3}\langle e_Q\rangle^2\alpha \omega^3 C_{fi} \delta_{SS'}|\langle n'^{2S'+1}L'_{J'}|r| n^{2S+1}L_J\rangle|^2 \frac{E_f}{M_i},
\end{equation}
where

\begin{equation}
  \langle e_Q\rangle=\frac{m_{\overline{q}}\;Q-m_q\;\overline{Q}}{m_q+m_{\overline{q}}}.
\end{equation}
Here $Q$ and $\overline{Q}$ are the charges of the quark and anti-quark in units of $|e|$, $m_q$ and $m_{\overline{q}}$ are the masses of the quark and anti-quark, $\alpha$ is the fine structure constant, $\omega$ is the final photon energy, $M_i$ is mass of the initial meson, and $E_f$ is the energy of the final state. Finally, the angular matrix element $C_{fi}$ is given by
\begin{equation}
 C_{fi}=max(L,L')\, (2J'+1)
                            \begin{Bmatrix}
                             L' & J' & S \\
                             J & L & 1 \\
                           \end{Bmatrix}^2. \nonumber
\end{equation}

This is the result of Ref.~\cite{godfrey 2004} except for our inclusion of the relativistic phase space factor $E_f/M_i$. The matrix elements $\langle n'^{2S'+1}L'_{J'}|r| n^{2S+1}L_J\rangle$ are obtained via numerical integration over the quark model wavefunctions obtained above.

Wavefunction shifts due to perturbative spin-dependent interactions were neglected in this computation, as with Refs. \cite{barnes 2005,close 2005}.
Results for E1 radiative transitions for $B$ and $B_s$ mesons are given in Tables \ref{d-1}-\ref{d-14}.

\subsection{M1 Radiative Transitions}

The M1 radiative partial widths are evaluated using the following expression\cite{godfrey 2016}
\begin{eqnarray}
\Gamma(n\;^{2S+1}L_J\rightarrow n'\;^{2S'+1}L_{J'}+\gamma) &=& \frac{\alpha}{3}\omega^3\, (2J'+1)\, \delta_{S,S'\pm1} \nonumber \\
  && \hspace{-2cm} \cdot \left|\frac{e_q}{m_q}\langle f|j_0(\frac{m_b}{m_q+m_b}k r)|i\rangle
+\frac{e_b}{m_b}\langle f|j_0(\frac{m_q}{m_q+m_b}k r)|i\rangle\right|^2,
\end{eqnarray}
where $\omega$ is the final photon energy and $j_0(x)$ is a spherical Bessel function. The definitions of the other parameters are same as in the E1 radiative transitions. We quote results for M1 radiative transitions for $B$ and $B_s$ mesons in Tables \ref{d-1}-\ref{d-14}.

\section{Results and Discussion}
\label{results}

\subsection{B Mesons}

Radiative decay rates, strong decay amplitudes, and partial widths for the 1P, 1D, 1F, 2S, 2P, 2D, and 3S $B$ mesons are given in
Tables \ref{d-1}-\ref{d-9}. In these tables $c=\cos\phi_{1P}$ or $\cos\phi_{2P}$, $s=\sin\phi_{1P}$ or $\sin\phi_{2P}$, $c_1=\cos\phi_{1D}$
or $\cos\phi_{2D}$, and $s_1=\sin\phi_{1D}$ or $\sin\phi_{2D}$. In the following we will use these results and those of Table \ref{B table} to interpret the recently discovered $B_1(5721)$, $B_2^*(5747)$, and $B_J(5970)$.

\subsubsection{$B_1(5721)$}

In 2016, the LHCb collaboration observed charged and neutral $B$ states, denoted as $B_1(5721)$\cite{LHCb 2016}. The spin-parity of these states was determined to be  $J^P=1^+$ while their masses are

$$m_{B_1(5721)^+}=5725.1\pm 1.8\pm 3.1\pm 0.17\pm 0.4 \ \textrm{MeV}$$
$$m_{B_1(5721)^0}=5727.7\pm 0.7\pm 1.4\pm 0.17\pm 0.4 \  \textrm{MeV}.$$
The measured decay widths are
$$\Gamma_{B_1(5721)^+}=29.1\pm 3.6\pm 4.3\ \textrm{MeV}$$
and
$$\Gamma_{B_1(5721)^0}=30.1\pm 1.5\pm 3.5 \ \textrm{MeV}.$$
Comparison of these mass to our calculated values $m_B(1P_1)=5755.12$ MeV and $m_B(1P'_1)=5739.2$ MeV implies that this state is one of the $1P_1$ or $1P_1'$ mesons. The ambiguity is resolved by comparison to the strong widths shown in Table IV, which shows
$\Gamma_{B(1P_1)}=16.4$ MeV and $\Gamma_{B(1P'_1)}=126.4$ MeV. Thus it is likely that the reported state is the narrow member of the $1P$ doublet, namely the  $B(1P_1)$.

\subsubsection{$B^*_2(5747)$}

The LHCb collaboration also observed spin partners of the $B_1(5721)$, dubbed $B^*_2(5747)$\cite{LHCb 2016}. This state has $J^P = 2^+$ and was observed in two charge modes with masses

$$m_{B^*_2(5747)^+}=5737.2\pm 0.72\pm 0.40\pm 0.17\ \textrm{MeV},$$
$$m_{B^*_2(5747)^0}=5739.44\pm 0.37\pm 0.33\pm 0.17\ \textrm{MeV}$$
and with decay widths of
$$\Gamma_{B^*_2(5747)^+}=23.6\pm 2.0\pm 2.1\ \textrm{MeV}$$
and
$$\Gamma_{B^*_2(5747)^0}=24.5\pm 1.0\pm 1.5\ \textrm{MeV}.$$

The measured properties of this state agree very well with the predicted mass and width of the $1{}^3P_2$ state (5767 MeV and 20.4 MeV, respectively).
In addition, we obtain the ratio

\begin{equation}
  \frac{\Gamma(B_2^*(5747)\rightarrow \pi B^*)}{\Gamma(B_2^*(5747)\rightarrow \pi B)}=1.002,
\end{equation}
which is also in good agreement with the experimental measurement\cite{LHCb 2016}:
$$\Gamma(B_2^*(5747)^+\rightarrow B^{*0}\pi^+)/\Gamma(B_2^*(5747)^+\rightarrow B^{0}\pi^+)=1.0\pm0.5\pm0.8$$
$$\Gamma(B_2^*(7)^0\rightarrow B^{*+}\pi^-)/\Gamma(B_2^*(5747)^0\rightarrow B^{+}\pi^-)=0.71\pm0.14\pm0.3 .$$

\subsubsection{$B_J(5840)$ and $B_J(5970)$}

Two "structures" labelled $B_J(5970)$ and $B_J(5840)$ have been reported by CDF\cite{CDF 2014} and LHCb\cite{lhcb2015}. 
Averaging the charge modes yields masses and widths of

\begin{equation}
M(B_J(5840)) = 5857 \pm 21 \ {\rm Mev} \qquad \Gamma(B_J(5840)) = 176 \pm 90\ {\rm MeV}
\end{equation}
and`
\begin{equation}
M(B_J(5970)) = 5968 \pm 7 \ {\rm Mev} \qquad \Gamma(B_J(5970)) = 72 \pm 23\  {\rm MeV}.
\end{equation}
The spin and parity of these states is unknown; however, the LHCb collaboration has suggested that the $B_J$ signals be identified with the $2^1S_0$ and $2^3S_1$ quark model bottom states. The measured masses and widths are difficult to fit with quark models. For example, the Godfrey-Isgur and ARM models of Ref. \cite{godfrey 2016} predict as $2S$ mass splitting of 39 and 30 MeV respectively, while we predict a splitting of 28 MeV. These are substantially less than the measured splitting $111 \pm 22$ MeV. This observation stands in contrast to the $n=1$ hyperfine splitting, where the measured difference is 45 MeV and the model predictions are 59 MeV (Godfrey-Isgur), 41 MeV (ARM), and 62 MeV (this work). 

In spite of this problem, the $B_J(5840)$ mass is reasonably close to model predictions for the $2^1S_0$ state, being 37 MeV below the Godfrey-Isgur result, 23 MeV above the ARM prediction, and 20 MeV above our prediction.  However, our predicted width for this state is 16 MeV, fairly low with respect to the measured width of 176 MeV (although with large error). In contrast, the width predicted in Ref. \cite{godfrey 2016} (95 MeV) is in good agreement with experiment. One concludes that identifying the $B_J(5840)$ with the $2^1S_0$ quark model state is reasonable. We note, however, that the $B\pi$ decay mode is reported as "possibly seen" by LHCb \cite{lhcb2015}, which would require dismissing the $2^1S_0$ identification if confirmed.

The quark model states closest to the $B_J(5970)$ in mass are the ground state $D$ waves, with masses near 6030 MeV, approximately 60 MeV too high. This is possibly an acceptable error since the $B$ and $B_s$ flavour sectors are rather poorly constrained.  The strong decay widths for these states, reported in Table \ref{tab:5}, range from 34 to 67 MeV. All of these match within errors with the measured width, with preference for the $1^3D_1$ or $1D_2'$ states. Again, the $B\pi$ decay is "possibly seen". If this is confirmed then the $1D_2'$ mode is disallowed. Overall, we slightly prefer the $1^3D_1$ identification of the $B_J(5970)$.

\subsubsection{$B$ meson decay properties}

\begin{table}[H]
\centering
\renewcommand{\arraystretch}{0.8}
\caption{Partial widths and branching ratios for strong, E1 and M1 decays of the 1S and 1P states of B mesons. E1 and M1 amplitudes are in units of $(GeV)^{-1}$ and strong decay amplitudes are in units of $(GeV)^{-1/2}$.}
\begin{tabular}{c c c c c c}
\hline\hline
State &  Mode & Photon energy & Amplitude    & $\Gamma_{thy}(ub,db)$ & B.R($ub$,$db$) \\
      &       & $(MeV)$       &              & $MeV$                        &($\%$) \\
\hline
$B(1 ^3S_1)$  &$B\gamma$      &    45.74     &$\langle 1^1S_0|j_0(kr\frac{m_{b,q}}{m_q+m_b})|1^3S_1\rangle=0.9930,0.9961$ & 0.0009,0.0003& 100 \\
\hline
$B(1 ^3P_0)$  &$B^*\gamma$    &   365.89     &$\langle 1^3S_1|r|1^3P_0\rangle=3.258$   & 0.575,0.175  & 0.4,0.12 \\
              &$ \pi B$       &              & $^1S_0=-0.392$             & 141.5  & $\sim$100 \\
              &total          &              &                            & 142.08,141.7  & 100 \\
\hline
$B(1 P_1)$    &$B\gamma$      &  441.51      &$\langle 1^1S_0|r|1^1P_1\rangle=2.965$   & 0.448,0.137  & 2.73,0.86 \\
              &$B^*\gamma$    &  399.07      &$\langle 1^3S_1|r|1^3P_1\rangle=3.258$   & 0.339,0.103  & 2.07,0.65 \\
              &$ \pi B^*$     &              & $^3S_1=-0.222 c+ 0.313 s$, & 15.62 & 96.7 \\
              &               &              & $^3D_1=-0.109 c- 0.077 s$  &        &   \\
              &total          &              &                            & 16.4,15.9 & 100 \\
\hline
$B(1 P_1^{'})$&$ B\gamma$     & 456.19       &$\langle 1^1S_0|r|1^1P_1\rangle=2.965$   &0.415,0.127   &  0.328,0.1  \\
              &$ B^*\gamma$   & 413.86       &$\langle 1^3S_1|r|1^3P_1\rangle=3.258$   &0.448,0.137   &  0.1,0.11  \\
              &$ \pi B^*$     &              & $^3S_1=+0.296 c+0.211 s$,  &125.53   &  $\sim$100 \\
              &               &              & $^3D_1=-0.082 c+0.115 s$   &        &    \\
              &total          &              &                            &126.4,125.8  & 100 \\
\hline
$B(1 ^3P_2)$  &$B^*\gamma$    & 402.6        & $\langle 1^3S_1|r|1^3P_2\rangle=3.258$  & 0.761,0.232  & 3.7,1.17    \\
              &$ \pi B$       &              & $^1D_2=+0.094$             & 9.77   & 48.7    \\
              &$\pi B^*$      &              & $^3D_2=-0.105$             & 9.79   & 48.8 \\
              &total          &              &                            &20.3,19.8 & 100         \\
\hline\hline
\end{tabular}
\label{d-1}
\end{table}

\begin{table}[H]
\centering
\renewcommand{\arraystretch}{0.8}
\caption{Partial widths and branching ratios for strong, E1 and M1 decays of the 1D states of B mesons (format as in Table \ref{d-1}).}
\label{tab:5}
\begin{tabular}{c c c c c c}
\hline\hline
State &  Mode & Photon energy & Amplitude    & $\Gamma_{thy}(ub,db)$ & B.R($ub$,$db$) \\
      &       & $(MeV)$       &              & $MeV$                        &($\%$) \\

\hline
$B(1 ^3D_1)$ & $B(1^3P_0)\gamma$&310.31   &$\langle 1^3P_0|r|1^3D_1\rangle=5.399$ & 0.65,0.199 & 0.9745,0.3002\\
             & $B(1^3P_2)\gamma$&272.91   &$\langle 1^3P_2|r|1^3D_1\rangle=5.399$ & 0.022,0.007 & 0.033,0.0106\\
                &$ \pi B$     &           & $^1P_1=+0.093$  & 23.22  &  34.9 \\
                &$\pi B^*$    &           & $^3P_1=+0.073$  & 12.49  &  18.78\\
                &$\eta B$     &           & $^1P_1=+0.065$  & 7.99   & 12.0 \\
                &$\eta B^*$   &           & $^3P_1=+0.050$  & 3.87   & 5.8   \\
                &$K B_s$      &           & $^1P_1=+0.095$  & 13.21  & 19.8  \\
                &$K B_s^*$    &           & $^3P_1=+0.068$  & 5.28   & 7.9  \\
                &total        &           &                 & 66.7,66.3  & 100 \\
\hline
$B(1 ^3D_3)$    & $B(1^3P_2)\gamma$&280.98&$\langle 1^3P_2|r|1^3D_3\rangle=5.399$ & 0.874,0.267 & 2.3245,0.7216\\
                &$ \pi B$     &           & $^1F_3=+0.076$  & 15.69  &  42\\
                &$\pi B^*$    &           & $^3F_3=-0.090$  & 19.49  & 52.2\\
                &$\eta B$     &           & $^1F_3=+0.017$  & 0.56   & 1.5\\
                &$\eta B^*$   &           & $^3F_3=-0.017$  & 0.44   & 1.17\\
                &$K B_s$      &           & $^1F_3=+0.015$  & 0.36   & 0.96\\
                &$K B_s^*$    &           & $^3F_3=-0.013$  & 0.20   & 0.53\\
                &total        &           &                 & 37.6,37.0  & 100 \\
\hline
$B(1 D_2)$      &$B(1^3P_2)\gamma$&278.63   &$\langle 1^3P_2|r|1^3D_2\rangle=5.399$ & 0.212,0.065 &0.6127,0.189 \\
                &$ \pi B^*$   &           & $^3P_2=+0.078c_{1}-0.095s_{1}$, &33.64  & 97.5 \\
                &             &           & $^3F_2=+0.092c_{1}+0.075s_{1}$  &       &  \\
                &$\eta B^*$   &           & $^3P_2=+0.055c_{1}-0.067s_{1}$, &0.74  & 2.1\\
                &             &           & $^3F_2=+0.017c_{1}+0.014s_{1}$  &       & \\
                &total        &           &                                 &34.6,34.4  & 100 \\
\hline
$B(1 D_2^{'})$  &$B(1^3P_2)\gamma$&279.53   &$\langle 1^3P_2|r|1^3D_2\rangle=5.399$ & 0.0011,0.0003 & 0.002,0.0005\\
                &$ \pi B^*$   &           & $^3P_2=-0.095c_{1}-0.077s_{1}$, &36.19  &  56.11\\
                &             &           & $^3F_2=+0.075c_{1}-0.092s_{1}$  &       &  \\
                &$\eta B^*$   &           & $^3P_2=-0.067c_{1}-0.055s_{1}$, &11.88   & 18.42 \\
                &             &           & $^3F_2=+0.014c_{1}-0.017s_{1}$  &       &  \\
                &$K B_s^*$    &           & $^3P_2=-0.092c_{1}-0.075s_{1}$, &16.43   &25.47  \\
                &             &           & $^3F_2=+0.010c_{1}-0.013s_{1}$  &       &  \\
                &total        &           &                                 &64.5,64.5  & 100        \\
\hline\hline
\end{tabular}
\label{d-2}
\end{table}

\begin{table}[H]
\centering
\renewcommand{\arraystretch}{0.8}
\caption{Partial widths and branching ratios for strong, E1 and M1 decays of the 1F states of B mesons (format as in Table \ref{d-1}).}
\begin{tabular}{c c c c c c}
\hline\hline
State &  Mode & Photon energy & Amplitude    & $\Gamma_{thy}(ub,db)$ & B.R($ub$,$db$) \\
      &       & $(MeV)$       &              & $MeV$                        &($\%$) \\
\hline
$B(1 ^3F_2)$    &$B(1^3D_1)\gamma$&232.21& $\langle 1^3D_1|r|1^3F_2\rangle=7.076$  &0.77,0.235& 1.08,0.34\\
                &$B(1D_2)\gamma$  &226.45& $\langle 1^3D_2|r|1^3F_2\rangle=7.076$  &0.71,0.217& 1.0,0.31\\
                &$B(1D'_2)\gamma$  &225.53& $\langle 1^3D_2|r|1^3F_2\rangle=7.076$  &0.0035,0.001& 0.005,0.001\\
                &$B(1^3D_3)\gamma$&224.16& $\langle 1^3D_3|r|1^3F_2\rangle=7.076$  &0.004,0.001& 0.006,0.001\\
                &$ \pi B$  &        & $^1D_2=+0.032$  &4.40  & 6.2\\
                &$\pi B^*$ &        & $^3D_2=+0.028$  &3.19  & 4.5\\
                &$\eta B$  &        & $^1D_2=+0.022$  &1.87  & 2.6\\
                &$\eta B^*$&        & $^3D_2=+0.021$  &1.46  & 2.1\\
                &$\rho B$  &        & $^3D_2=+0.054$  &8.12  & 11.6\\
                &$\rho B^*$&        & $^1D_2=+0.045$,$^5D_2=-0.034$,$^5G_2=-0.095$ &28.92  & 41\\
                &$\omega B$&        & $^3D_2=+0.032$  &2.77  & 3.9\\
                &$\omega B^*$&      & $^1D_2=+0.026$,$^5D_2=-0.020$,$^5G_2=-0.052$ &8.65   & 12.3\\
                &$ \pi B(2^1S_0)$  && $^1D_2=-0.053$  &2.14  & 3.0\\
                &$\pi B(2^3S_1)$ &  & $^3D_2=-0.041$  &1.11  & 1.6\\
                &$K B_s$   &        & $^1D_2=+0.035$  &3.98  & 5.6\\
                &$K B_s^*$ &        & $^3D_2=+0.032$  &2.92  & 4.1\\
                &total     &        &                 & 71,70& 100  \\
\hline
$B(1 ^3F_4)$    &$B(1^3D_3)\gamma$&217.33& $\langle 1^3D_3|r|1^3F_4\rangle=7.076$  &0.753,0.23& 0.7591,0.2323\\
                &$ \pi B$  &        & $^1G_4=+0.046$  &9.18  &  9.26\\
                &$\pi B^*$ &        & $^3G_4=-0.054$  &12.42 & 12.53\\
                &$\eta B$  &        & $^1G_4=+0.015$  &0.82  & 0.82\\
                &$\eta B^*$&        & $^3G_4=-0.017$  &1.01  & 1.01\\
                &$\rho B$  &        & $^3G_4=-0.057$  &8.92  & 9\\
                &$\rho B^*$&        & $^5D_4=-0.134$,$^1G_4=+0.023$,$^5G_4=-0.045$ &46.49 & 46.9\\
                &$\omega B$&        & $^3G_4=-0.032$  &2.67  & 2.69\\
                &$\omega B^*$&      & $^5D_4=-0.077$,$^1G_4=+0.012$,$^5G_4=-0.024$ &15.03 & 15.16\\
                &$K B_s$   &        & $^1G_4=+0.017$  &0.93  & 0.94\\
                &$K B_s^*$ &        & $^3G_4=-0.019$  &0.98  & 0.99\\
                &total     &        &                 &99.2,99 & 100                \\
\hline\hline
\end{tabular}
\label{d-3}
\end{table}
\begin{table}[H]
\tabcolsep=1pt\fontsize{10}{10}\selectfont
\centering
\renewcommand{\arraystretch}{0.8}
\caption{Partial widths and branching ratios for strong, E1 and M1 decays of the 2S states of B mesons (format as in Table \ref{d-1}).}
\begin{tabular}{c c c c c c}
\hline\hline
State &  Mode & Photon energy & Amplitude    & $\Gamma_{thy}(ub,db)$ & B.R($ub$,$db$) \\
      &       & $(MeV)$       &              & $MeV$                        &($\%$) \\

\hline
$B(2 ^1S_0)$    &$B^*\gamma$    &526.18 & $\langle 1^3S_1|j_0(kr\frac{m_{b,q}}{m_q+m_b})|2^1S_0\rangle=0.1985,-0.0695$  &0.181,0.0421 &1.1603, 0.2752\\
                &$B(1P_1)\gamma$&136.49 & $\langle 1^1P_1|r|2^1S_0\rangle=-4.4898$                                      &0.096,0.0294& 0.6154,0.1922\\
                &$B(1P_1')\gamma$&120.92& $\langle 1^1P_1|r|2^1S_0\rangle=-4.4898$                                      &0.057,0.0173& 0.3654,0.1131\\
                &$ \pi B^*$ &       & $^3P_0=-0.100$ & 15.25 & $\sim$100 \\
                & total     &       &                & 15.6,15.3 & 100           \\
\hline
$B(2 ^3S_1)$    &$B\gamma$      &592.91 & $\langle 1^1S_0|j_0(kr\frac{m_{b,q}}{m_q+m_b})|2^3S_1\rangle=0.2770,0.0700$  &0.161,0.0423  & 2.205,0.613\\
                &$B(2^1S_0)\gamma$&27.98& $\langle 2^1S_0|j_0(kr\frac{m_{b,q}}{m_q+m_b})|2^3S_1\rangle=0.9897,0.9949$  &0.0002,0.0001 &0.003,0.001\\
                &$B(1^3P_2)\gamma$&160.13 & $\langle 1^3P_2|r|2^3S_1\rangle=-4.3370$                                   &0.148,0.045   &2.027,0.652\\
                &$B(1P_1)\gamma$&163.82 & $\langle 1^3P_1|r|2^3S_1\rangle=-4.3370$                                     &0.043,0.0133  &0.589,0.193\\
                &$B(1P_1')\gamma$&148.33& $\langle 1^3P_1|r|2^3S_1\rangle=-4.3370$                                     &0.038,0.012   &0.521,0.174\\
                &$B(1^3P_0)\gamma$&198.27 & $\langle 1^3P_0|r|2^3S_1\rangle=-4.3370$                                   &0.056,0.017   &0.767,0.246\\
                &$ \pi B$   &       & $^1P_1=-0.017$ & 0.56  &7.67,8.12 \\
                &$\pi B^*$  &       & $^3P_1=-0.052$ & 4.51  &61.78,65.36\\
                &$\eta B$   &       & $^1P_1=-0.025$ & 0.64  &8.77,9.28\\
                &$\eta B^*$ &       & $^3P_1=-0.034$ & 0.7   &9.59,10.14\\
                &$K B_s$    &       & $^1P_1=-0.025$ & 0.40  &5.48,5.80\\
                &total      &       &                & 7.3,6.9  &100\\
\hline\hline
\end{tabular}
\label{d-4}
\end{table}

\begin{table}[H]
\centering
\renewcommand{\arraystretch}{0.8}
\caption{Partial widths and branching ratios for strong, E1 and M1 decays of the 2P states of B mesons (format as in Table \ref{d-1}).}
\begin{tabular}{c c c c c c}
\hline\hline
State &  Mode & Photon energy & Amplitude    & $\Gamma_{thy}(ub,db)$ & B.R($ub$,$db$) \\
      &       & $(MeV)$       &              & $MeV$                        &($\%$) \\

\hline
$B(2 ^3P_0)$  &$B^*\gamma$        &750.80   &$\langle 1^3S_1|r|2^3P_0\rangle=0.5734$  & 0.144,0.044 &0.159,0.049 \\
              &$B(2^3S_1)\gamma$  &219.24   &$\langle 2^3S_1|r|2^3P_0\rangle=5.21926$ & 0.327,0.0998 &0.361,0.111 \\
              &$B(1^3D_1)\gamma$  &105.36   &$\langle 1^3D_1|r|2^3P_0\rangle=-4.5973$ & 0.057,0.018 &0.063,0.02 \\
              &$ \pi B$  &       & $^1S_0=-0.066$ & 14.72 & 16.2 \\
              &$ \eta B$ &       & $^1S_0=-0.058$ & 8.88  &9.8 \\
              &$ \rho B^*$&      & $^1S_0=+0.138$,$^5D_0=-0.106$ & 25.99  & 28.69\\
              &$ \omega B^*$&    & $^1S_0=+0.090$,$^5D_0=-0.048$ & 7.61  & 8.4\\
              &$ \pi B(2^1S_0)$& & $^1S_0=-0.175$ & 9.43  & 10.41\\
              &$ K B_s$  &       & $^1S_0=-0.103$ & 23.45 & 25.88 \\
              &total     &       &                & 90.6,90.2 & 100\\
\hline
$B(2 ^3P_2)$  &$B^*\gamma$        &804.09   &$\langle 1^3S_1|r|2^3P_2\rangle=0.5734$  & 0.176,0.054  &0.129,0.04 \\
              &$B(2^3S_1)\gamma$  &277.76   &$\langle 2^3S_1|r|2^3P_2\rangle=5.21926$ & 0.659,0.2    &0.483,0.147 \\
              &$B(1^3D_1)\gamma$  &165      &$\langle 1^3D_1|r|2^3P_2\rangle=-4.5973$ & 0.0022,0.0007&0.002,0.001 \\
              &$B(1^3D_3)\gamma$  &156.85   &$\langle 1^3D_3|r|2^3P_2\rangle=-4.5973$ & 0.158,0.048  &0.116,0.035 \\
              &$ \pi B$  &       & $^1D_2=-0.070$ & 18.58 & 13.6 \\
              &$\pi B^*$ &       & $^3D_2=+0.074$ & 19.25 &14.1\\
              &$\eta B$  &       & $^1D_2=-0.017$ & 0.89  & 0.65\\
              &$\eta B^*$&       & $^3D_2=+0.013$ & 0.47  &0.34\\
              &$\rho B$  &       & $^3D_2=+0.116$ & 28.62 & 21\\
              &$\rho B^*$&       & $^5S_2=+0.054$,$^1D_2=+0.053$,$^5D_2=-0.014$ & 41.78 & 30.6 \\
              &$\omega B$&       & $^3D_2=+0.067$ & 9.16  & 6.7\\
              &$\omega B^*$&     & $^5S_2=+0.042$,$^1D_2=+0.030$,$^5D_2=-0.078$ & 13.76 & 10.1\\
              &$\pi B(2^1S_0)$&  & $^1D_2=-0.035$ & 0.6   & 0.44 \\
              &$\pi B(2^3S_1)$ & & $^3D_2=+0.038$ & 0.6   &0.44\\
              &$K B_s$   &       & $^1D_2=-0.022$ & 1.23  & 0.9\\
              &$K B_s^*$ &       & $^3D_2=+0.016$ & 0.61  & 0.45\\
              &total     &       &                & 136.5,136.2 & 100  \\
\hline\hline
\end{tabular}
\label{d-5}
\end{table}

\begin{table}[H]
\centering
\renewcommand{\arraystretch}{0.8}
\caption{Partial widths and branching ratios for strong, E1 and M1 decays of the 2P states(continued) of B mesons (format as in Table \ref{d-1}).}
\begin{tabular}{c c c c c c}
\hline\hline
State &  Mode & Photon energy & Amplitude    & $\Gamma_{thy}(ub,db)$ & B.R($ub$,$db$) \\
      &       & $(MeV)$       &              & $MeV$                        &($\%$) \\

\hline
$B(2 P_1)$    &$B\gamma$          &818.99   &$\langle 1^1S_0|r|2^1P_1\rangle=0.670513$  & 0.112,0.0419 & 0.084,0.031\\
              &$B^*\gamma$        &779.46   &$\langle 1^3S_1|r|2^3P_1\rangle=0.5734$    & 0.073,0.0224 & 0.054,0.017\\
              &$B(2^1S_0)\gamma$  &277.54   &$\langle 2^1S_0|r|2^1P_1\rangle=4.84082$   & 0.307,0.094  & 0.229,0.07\\
              &$B(2^3S_1)\gamma$  &250.72   &$\langle 2^3S_1|r|2^3P_1\rangle=5.21926$   & 0.223,0.068  & 0.166,0.051\\
              &$ \pi B^*$&       & $^3S_1=-0.036c+0.051s$  &25.34 & 18.9     \\
              &          &       & $^3D_1=-0.071c-0.051s$  &      &     \\
              &$\eta B^*$&       & $^3S_1=-0.032c+0.046s$  &0.38  & 0.28     \\
              &          &       & $^3D_1=-0.010c-0.007s$  &      &     \\
              &$\rho B$  &       & $^3S_1=-0.008c+0.011s$  &36.67  & 27.4     \\
              &          &       & $^3D_1=-0.116c-0.082s$  &      &     \\
              &$\omega B$&       & $^3S_1=-0.009c+0.012s$  &11.24   & 8.4     \\
              &          &       & $^3D_1=-0.065c-0.046s$  &      &     \\
              &$\rho B^*$&       & $^3S_1=0.069c$,$^3D_1=0.154c$,$^5D_1=-0.188s$  &45.38  &  33.9    \\
              &$\omega B^*$&     & $^3S_1=0.049c$,$^3D_1=0.084c$,$^5D_1=-0.102s$  &13.77  &  10.3    \\
              &$K B_s^*$ &       & $^3S_1=-0.058c+0.082s$  &0.51  & 0.38     \\
              &          &       & $^3D_1=-0.013c-0.009s$  &      &     \\
              &total     &       &                         &134,133.5 & 100  \\
\hline
$B(2 P_1^{'})$ &$B\gamma$         &830.99   &$\langle 1^1S_0|r|2^1P_1\rangle=0.670513$  & 0.121,0.037 & 0.138,0.042\\
              &$B^*\gamma$        &791.54   &$\langle 1^3S_1|r|2^3P_1\rangle=0.5734$    & 0.091,0.028 &0.104,0.032 \\
              &$B(2^1S_0)\gamma$  &290.74   &$\langle 2^1S_0|r|2^1P_1\rangle=4.84082$ & 0.297,0.09 & 0.338,0.103\\
              &$B(2^3S_1)\gamma$  &263.99   &$\langle 2^3S_1|r|2^3P_1\rangle=5.21926$   & 0.308,0.094 & 0.351,0.108\\
              &$ \pi B^*$&      & $^3S_1=+0.046c+0.033s$  &10.89 & 12.4   \\
               &          &      & $^3D_1=-0.053c+0.074s$  &      &     \\
               &$\eta B^*$&      & $^3S_1=+0.044c+0.031s$  &7.87  & 8.96     \\
               &          &      & $^3D_1=-0.008c+0.011s$  &      &     \\
               &$\rho B$  &      & $^3S_1=+0.008c+0.005s$  &0.19 & 0.22     \\
               &          &      & $^3D_1=-0.085c+0.120s$  &      &     \\
               &$\omega B$&      & $^3S_1=+0.0002c+0.0005s$&0.001 & 0.001       \\
               &          &      & $^3D_1=-0.048c+0.068s$  &      &     \\
               &$\rho B^*$&      & $^3S_1=-0.102s$,$^3D_1=-0.135s$,$^5D_1=-0.166c$ &36.59 &  41.67    \\
               &$\omega B^*$&    & $^3S_1=-0.069s$,$^3D_1=-0.072s$,$^5D_1=-0.088c$ &10.49 &  11.95    \\
               &$K B_s^*$ &      & $^3S_1=+0.080c+0.056s$  &20.99  & 23.91     \\
               &          &      & $^3D_1=-0.011c+0.015s$  &      &     \\
               &total     &      &                         &87.8,87.3 & 100  \\
\hline\hline
\end{tabular}
\label{d-6}
\end{table}
\begin{table}[H]
\centering
\renewcommand{\arraystretch}{0.8}
\caption{Partial widths and branching ratios for strong, E1 and M1 decays of the 2D states of B mesons (format as in Table \ref{d-1}).}
\begin{tabular}{c c c c c c}
\hline\hline
State &  Mode & Photon energy & Amplitude    & $\Gamma_{thy}(ub,db)$ & B.R($ub$,$db$) \\
      &       & $(MeV)$       &              & $MeV$                        &($\%$) \\
\hline
$B(2 ^3D_1)$    &$B(1^3P_0)\gamma$  &644.45   &$\langle 1^3P_0|r|2^3D_1\rangle=0.491304$ & 0.046,0.014 & 0.0601,0.0183\\
                &$B(1^3P_2)\gamma$  &609.16   &$\langle 1^3P_2|r|2^3D_1\rangle=0.491304$ & 0.002,0.0006 &0.0026,0.0008 \\
                &$ \pi B$    &        & $^1P_1=-0.016$  & 1.43 & 1.87\\
                &$\pi B^*$   &        & $^3P_1=-0.013$  & 0.83 & 1.08 \\
                &$\eta B$    &        & $^1P_1=-0.019$  & 1.79 & 2.34\\
                &$\eta B^*$  &        & $^3P_1=-0.015$  & 1.02 & 1.33\\
                &$\rho B$    &        & $^3P_1=-0.031$  & 3.74 & 4.89\\
                &$\rho B^*$  &        & $^1P_1=+0.026$,$^5P_1=-0.012$,$^5F_1=-0.102$ & 40.11 & 52.46 \\
                &$\omega B$  &        & $^3P_1=-0.018$  & 1.25 & 1.63\\
                &$\omega B^*$&        & $^1P_1=+0.015$,$^5P_1=-0.007$,$^5F_1=-0.061$ & 14.12 & 18.46\\
                &$K B_s$     &        & $^1P_1=-0.041$  & 6.99 & 9.14\\
                &$K B_s^*$   &        & $^3P_1=-0.032$  & 3.80 & 4.97\\
                &$K^* B_s$   &        & $^3P_1=-0.022$  & 1.10 & 1.43\\
                &$K^* B_s^*$ &        & $^1P_1=+0.010$,$^5P_1=-0.005$,$^5F_1=-0.006$, & 0.28 & 0.36\\
                &total       &        &                 & 76.5,76.4 & 100 \\
\hline
$B(2 ^3D_3)$    &$B(1^3P_2)\gamma$    &619.16   &$\langle 1^3P_2|r|2^3D_3\rangle=0.491304$ & 0.073,0.0224 &0.063,0.019 \\
                &$ \pi B$    &        & $^1F_3=+0.067$  & 24.78 & 21.3 \\
                &$\pi B^*$   &        & $^3F_3=-0.076$  & 29.40 & 25.3\\
                &$\eta B$    &        & $^1F_3=+0.024$  & 2.81  & 2.42\\
                &$\eta B^*$  &        & $^3F_3=-0.026$  & 2.98  & 2.56\\
                &$\rho B$    &        & $^3F_3=-0.028$  & 3.19  & 2.75\\
                &$\rho B^*$  &        & $^5P_3=-0.072$,$^1F_3=+0.023$,$^5F_3=-0.050$  & 30.38 & 26.14\\
                &$\omega B$  &        & $^3F_3=-0.017$  & 1.22  & 1.06 \\
                &$\omega B^*$&        & $^5P_3=-0.041$,$^1F_3=+0.014$,$^5F_3=-0.030$ & 10.25 & 8.82\\
                &$\pi B(2^1S_0)$ &    & $^1F_3=+0.018$  & 0.46 & 0.4 \\
                &$\pi B(2^3S_1)$ &    & $^3F_3=-0.023$  & 0.67 & 0.58\\
                &$K B_s$     &        & $^1F_3=+0.031$  & 4.00  & 3.44\\
                &$K B_s^*$   &        & $^3F_3=-0.032$  & 3.98  & 3.43\\
                &$K^* B_s$   &        & $^3F_3=-0.0005$ & $7.20\times 10^{-4}$ & 0.001\\
                &$K^* B_s^*$ &        & $^5P_3=-0.032$,$^1F_3=+0.002$,$^5F_3=-0.003$  & 1.97  & 1.7\\
                &total       &        &                 & 116.2,116.1  & 100\\
\hline\hline
\end{tabular}
\label{d-7}
\end{table}

\begin{table}[H]
\centering
\renewcommand{\arraystretch}{0.8}
\caption{Partial widths and branching ratios for strong, E1 and M1 decays of the 2D states(continued) of B mesons (format as in Table \ref{d-1}).}
\begin{tabular}{c c c c c c}
\hline\hline
State &  Mode & Photon energy & Amplitude    & $\Gamma_{thy}(ub,db)$ & B.R($ub$,$db$) \\
      &       & $(MeV)$       &              & $MeV$                        &($\%$) \\
\hline
$B(2 D_2)$      &$B(1^3P_2)\gamma$    &615.82   &$\langle 1^3P_2|r|2^3D_2\rangle=0.491304$ & 0.018,0.005 & 0.016,0.005\\
                &$B(2^3P_2)\gamma$    &198.9    &$\langle 2^3P_2|r|2^3D_2\rangle=7.13132$  & 0.137,0.042 & 0.125,0.038\\
                &$ \pi B^*$ &       & $^3P_2=+0.013c_1-0.016s_1$, &50.91 & 46.6 \\
                &           &       & $^3F_2=+0.078c_1+0.064s_1$  &     &   \\
                &$\eta B^*$ &       & $^3P_2=+0.016c_1-0.020s_1$, &5.1 & 4.67\\
                &           &       & $^3F_2=+0.026c_1+0.021s_1$  &     &  \\
                &$\rho B$   &       & $^3P_2=+0.033c_1-0.041s_1$, &6.11& 5.6 \\
                &           &       & $^3F_2=+0.030c_1+0.025s_1$  &     &  \\
                &$\omega B$ &       & $^3P_2=+0.019c_1-0.024s_1$, &2.31 & 2.1 \\
                &           &       & $^3F_2=+0.019c_1+0.015s_1$  &     &  \\
                &$\rho B^*$ &       & $^3P_2=-0.051c_1$,$^3F_2=-0.068c_1$, &26.55&24.3  \\
                &           &       & $^5P_2=+0.036s_1$,$^5F_2=+0.078s_1$  &     &  \\
                &$\omega B^*$ &     & $^3P_2=-0.029c_1$,$^3F_2=-0.041c_1$, &9.25& 8.5 \\
                &           &       & $^5P_2=+0.021s_1$,$^5F_2=+0.047s_1$  &     &  \\
                &$ \pi B(2^3S_1)$&  & $^3P_2=+0.009c_1-0.104s_1$, &1.17 & 1.1 \\
                &           &       & $^3F_2=+0.024c_1+0.019s_1$  &     &   \\
                &$ K B_s^*$ &       & $^3P_2=+0.034c_1-0.041s_1$, &6.85 & 6.3 \\
                &           &       & $^3F_2=+0.033c_1+0.027s_1$  &     &   \\
                &$ K^* B_s$ &       & $^3P_2=+0.025c_1-0.030s_1$, &0.002 & 0.002 \\
                &           &       & $^3F_2=+0.0006c_1+0.0005s_1$  &     &   \\
                &$K^* B_s^*$&       & $^3P_2=-0.022c_1$,$^3F_2=-0.004c_1$, &0.76& 0.7 \\
                &           &       & $^5P_2=+0.016s_1$,$^5F_2=+0.005s_1$  &     &  \\
                &total      &       &                             &109.2,109.1& 100   \\
\hline\hline
\end{tabular}
\label{d-8}
\end{table}
\begin{table}[H]
\centering
\renewcommand{\arraystretch}{0.8}
\caption{Partial widths and branching ratios for strong, E1 and M1 decays of the 2D states(continued) of B mesons (format as in Table \ref{d-1}).}
\begin{tabular}{c c c c c c}
\hline\hline
State &  Mode & Photon energy & Amplitude    & $\Gamma_{thy}(ub,db)$ & B.R($ub$,$db$) \\
      &       & $(MeV)$       &              & $MeV$                        &($\%$) \\
\hline
$B(2 D_2^{'})$  &$B(1^3P_2)\gamma$    &616.27   &$\langle 1^3P_2|r|2^3D_2\rangle=0.491304$ & 0.0001,0.00003 & 0.0001,0.00004\\
                &$B(2^3P_2)\gamma$    &199.38   &$\langle 2^3P_2|r|2^3D_2\rangle=7.13132$  & 0.001,0.0002 & 0.0014,0.00028\\
                &$ \pi B^*$ &       & $^3P_2=-0.016c_1-0.013s_1$, &2.08 & 2.88\\
                &           &       & $^3F_2=+0.064c_1-0.078s_1$  &      &  \\
                &$\eta B^*$ &       & $^3P_2=-0.020c_1-0.016s_1$, &2.87  & 3.97 \\
                &           &       & $^3F_2=+0.021c_1-0.026s_1$  &      &   \\
                &$\rho B$   &       & $^3P_2=-0.041c_1-0.033s_1$, &11.23  & 15.53 \\
                &           &       & $^3F_2=+0.024c_1-0.030s_1$  &      &  \\
                &$\omega B$ &       & $^3P_2=-0.024c_1-0.019s_1$, &3.76  & 5.2 \\
                &           &       & $^3F_2=+0.015c_1-0.018s_1$  &      &   \\
                &$\rho B^*$ &       & $^3P_2=+0.051s_1$,$^3F_2=+0.069s_1$, &27.2  & 37.62 \\
                &           &       & $^5P_2=+0.036c_1$,$^5F_2=+0.080c_1$  &      &  \\
                &$\omega B^*$ &     & $^3P_2=+0.029s_1$,$^3F_2=+0.042s_1$, &9.51  & 13.15 \\
                &           &       & $^5P_2=+0.021c_1$,$^5F_2=+0.048c_1$  &      &  \\
                &$ \pi B(2^3S_1)$&  & $^3P_2=-0.011c_1-0.009s_1$, &0.25 & 0.35 \\
                &           &       & $^3F_2=+0.019c_1-0.024s_1$  &      &  \\
                &$ K^* B_s$ &       & $^3P_2=-0.030c_1-0.025s_1$, &3.64 & 5.03\\
                &           &       & $^3F_2=+0.0006c_1-0.0007s_1$  &      &  \\
                &$ K B_s^*$ &       & $^3P_2=-0.042c_1-0.034s_1$, &11.12 & 15.38\\
                &           &       & $^3F_2=+0.027c_1-0.032s_1$  &      &  \\
                &$K^* B_s^*$&       & $^3P_2=+0.021s_1$,$^3F_2=+0.004s_1$, &0.62  & 0.86 \\
                &           &       & $^5P_2=+0.015c_1$,$^5F_2=+0.005c_1$  &      &  \\
                &total      &       &                             &72.3,72.3 & 100         \\
\hline\hline
\end{tabular}
\label{d-8}
\end{table}
\begin{table}[H]
\tabcolsep=1pt\fontsize{10}{10}\selectfont
\centering
\renewcommand{\arraystretch}{0.8}
\caption{Partial widths and branching ratios for strong, E1 and M1 decays of the 3S states of B mesons (format as in Table \ref{d-1}).}
\begin{tabular}{c c c c c c}
\hline\hline
State &  Mode & Photon energy & Amplitude    & $\Gamma_{thy}(ub,db)$ & B.R($ub$,$db$) \\
      &       & $(MeV)$       &              & $MeV$                        &($\%$) \\

\hline
$B(3 ^1S_0)$ &$B^*\gamma$       &889.07 & $\langle 1^3S_1|j_0(kr\frac{m_{b,q}}{m_q+m_b})|3^1S_0\rangle=0.17019,-0.03307$  &0.633,0.152 & 2.38,0.587\\
             &$B(2^3S_1)\gamma$ &370.98 & $\langle 2^3S_1|j_0(kr\frac{m_{b,q}}{m_q+m_b})|3^1S_0\rangle=0.24297,-0.05789$  &0.094,0.022 & 0.353,0.085\\
             &$B(1P_1)\gamma$   &524.83 & $\langle 1^1P_1|r|3^1S_0\rangle=0.195518$  &0.01,0.003 & 0.038,0.012\\
             &$B(1P'_1)\gamma$  &510.28 & $\langle 1^1P_1|r|3^1S_0\rangle=0.195518$  &0.008,0.002 & 0.03,0.008\\
             &$B(2P_1)\gamma$   &125.31 & $\langle 2^1P_1|r|3^1S_0\rangle=-6.90827$  &0.177,0.054 & 0.665,0.208\\
             &$B(2P'_1)\gamma$  &111.72 & $\langle 2^1P_1|r|3^1S_0\rangle=-6.90827$  &0.106,0.032 & 0.398,0.124\\
             & $\pi B^*$  &      & $^3P_0=+0.027$ & 3.12 & 11.73 \\
             & $\eta B^*$ &      & $^3P_0=+0.008$ & 0.23 & 0.88 \\
             & $\rho B$   &      & $^3P_0=+0.022$ & 1.49 & 5.6 \\
             & $\rho B^*$ &      & $^3P_0=+0.022$ & 1.23 & 4.62 \\
             & $\omega B$ &      & $^3P_0=+0.010$ & 0.30 & 1.13 \\
             & $\omega B^*$&     & $^3P_0=+0.018$ & 0.85 & 3.2 \\
             & $\pi B(2 ^3S_1)$& & $^3P_0=-0.053$ & 2.12 & 7.97 \\
             & $\pi B(1 ^3P_0)$& & $^1S_0=+0.086$ & 13.46 & 50.6 \\
             & $\pi B(1 ^3P_2)$& & $^5D_0=-0.033$ & 1.65 & 6.2 \\
             & $K B_s^*$  &      & $^3P_0=+0.016$ & 0.76 & 2.86 \\
             & $K^* B_s$  &      & $^3P_0=+0.009$ & 0.08 & 0.3 \\
             & $K B_s(1^3P_0)$&  & $^1S_0=-0.027$ & 0.33 & 1.24 \\
             & total      &      &                &26.6,26& 100 \\
\hline
$B(3 ^3S_1)$ &$B\gamma$         &945.03 & $\langle 1^1S_0|j_0(kr\frac{m_{b,q}}{m_q+m_b})|3^3S_1\rangle=0.19328,0.03665$  &0.319,0.0828 & 0.979,0.258\\
             &$B(2^1S_0)\gamma$ &416.18 & $\langle 2^1S_0|j_0(kr\frac{m_{b,q}}{m_q+m_b})|3^3S_1\rangle=0.27629,0.0587$   &0.056,0.0145 & 0.172,0.045\\
             &$B(3^1S_0)\gamma$ &20.20  & $\langle 3^1S_0|j_0(kr\frac{m_{b,q}}{m_q+m_b})|3^3S_1\rangle=0.99,0.99549$     &0.0001,0.00002 & 0.0003,0.0001\\
             &$B(1^3P_2)\gamma$   &539.89 & $\langle 1^3P_2|r|3^3S_1\rangle=0.06559$  &0.0012,0.0004 & 0.004,0.001\\
             &$B(1P_1)\gamma$  &543.35 & $\langle 1^3P_1|r|3^3S_1\rangle=0.06559$     &0.0003,0.0001 & 0.001,0.0003\\
             &$B(1P'_1)\gamma$   &528.84 & $\langle 1^3P_1|r|3^3S_1\rangle=0.06559$  &0.00037,0.0001 & 0.001,0.0003\\
             &$B(1^3P_0)\gamma$  &575.60 & $\langle 1^3P_0|r|3^3S_1\rangle=0.06559$  &0.0003,0.0001 & 0.001,0.0003\\
             &$B(2^3P_2)\gamma$   &117.45 & $\langle 2^3P_2|r|3^3S_1\rangle=-6.69759$&0.14,0.0428 & 0.429,0.1333\\
             &$B(2P_1)\gamma$  &145.10 & $\langle 2^3P_1|r|3^3S_1\rangle=-6.69759$  &0.072,0.0221 & 0.221,0.0688\\
             &$B(2P'_1)\gamma$   &131.56 & $\langle 2^3P_1|r|3^3S_1\rangle=-6.69759$  &0.064,0.0196 &0.196,0.0611 \\
             &$B(2^3P_0)\gamma$  &176.99 & $\langle 2^3P_0|r|3^3S_1\rangle=-6.69759$  &0.095,0.029 & 0.291,0.0903\\
             & $\pi B$    &      & $^1P_1=+0.033$ & 5.16  & 15.83,16.07\\
             &$\pi B^*$   &      & $^3P_1=+0.034$ & 5.19  & 15.92,16.17\\
             &$\eta B$    &      & $^1P_1=+0.003$ & 0.05  & 0.15,0.16\\
             &$\eta B^*$  &      & $^3P_1=+0.002$ & 0.01  & 0.03,0.03\\
             &$\rho B$    &      & $^3P_1=+0.035$ & 3.98  & 12.21,12.4\\
             &$\rho B^*$  &      & $^1P_1=-0.005$, $^5P_1=+0.023$ & 1.56 & 4.79,4.86\\
             &$\omega B$  &      & $^3P_1=+0.018$ & 1.10  & 3.37,3.43\\
             &$\omega B^*$&      & $^1P_1=-0.002$, $^5P_1=+0.008$  & 0.20 & 0.61,0.62\\
             & $\pi B(2^1S_0)$&  & $^1P_1=+0.014$ & 0.2  & 0.61,0.62\\
             & $\pi B(2^3S_1)$&  & $^3P_1=-0.030$ & 0.76  & 2.33,2.37\\
             & $\pi B(1P_1)$&    & $^3S_1=+0.072c+0.052s$ & 13.01  & 39.91,40.53\\
             &             &     & $^3D_1=+0.008c-0.010s$ &        & \\
             & $\pi B(1P'_1)$&   & $^3S_1=-0.052c+0.073s$ & 0.15  & 0.46,0.47\\
             &             &     & $^3D_1=+0.007c+0.006s$ &        & \\
             & $\pi B(1^3P_2)$&  & $^5D_1=-0.014$         & 0.31  &0.95,0.97 \\
             &$K B_s$     &      & $^1P_1=+0.001$ & 0.01 & 0.03,0.03\\
             &$K B_s^*$   &      & $^3P_1=+0.007$ & 0.17 & 0.52,0.53\\
             &total       &      &                & 32.6,32.1& 100          \\
\hline\hline
\end{tabular}
\label{d-9}
\end{table}

\subsection{$B_s$ Mesons}

Strong decay amplitudes, radiative transitions, and partial widths for the  1P, 1D, 1F, 2S, 2P, and 2D  $B_s$ mesons are given in
Tables \ref{d-10}-\ref{d-14}. In these tables $c=\cos\phi_{1P}$ or $\cos\phi_{2P}$, $s=\sin\phi_{1P}$ or $\sin\phi_{2P}$, $c_1=\cos\phi_{1D}$
or $\cos\phi_{2D}$, $s_1=\sin\phi_{1D}$ or $\sin\phi_{2D}$. These results are used to interpret the observed mesons $B_{s1}(5830)$ and $B_{s2}^*(5840)$.

\subsubsection{$B_{s1}(5830)$}

The CDF collaboration observed a  bottom-strange state, called the $B_{s1}(5830)$, with $J^P=1^+$ and mass\cite{CDF 2014}:

$$m_{B_{s1}(5830)}=5828.3\pm 0.1\pm 0.2\pm 0.4\ \textrm{MeV}.$$
The width of this state was determined to be
$$\Gamma_{B_{s1}(5830)}=0.5\pm0.3\pm 0.3 \ \textrm{MeV}.$$

Evidently this state is one of the $1{}P_1$ doublet mesons, which have predicted masses near 5800 MeV. These masses are just above $BK$ threshold and therefore small widths are predicted for the narrow $1P_1$ and (nominally) wide $1P_1'$ states. Because of this no further identification can be made.

\subsubsection{$B^*_{s2}(5840)$}

The CDF collaboration has also reported the $B^*_{s2}(5840)$ \cite{CDF 2014}. This is a  $J^P=2^+$ state of mass
$$m_{B^*_{s2}(5840)}=5839.7\pm 0.1\pm 0.1\pm 0.2 \ \textrm{MeV}$$
and width
$$\Gamma_{B^*_{s2}(5840)}=1.4\pm 0.4\pm 0.2\ \textrm{MeV}.$$

These values compare favourably with the predicted mass $m_{B_s}(1{}^3P_2)=5821.53$ MeV and width $\Gamma_{B_s}(1{}^3P_2)=1.9$ MeV.  In addition, we also obtain the ratio
\begin{equation}
  \frac{\Gamma(B_{s2}^*(5840)\rightarrow K B^*)}{\Gamma(B_{s2}^*(5840)\rightarrow K B)}=0.084,
\end{equation}
which is also in good agreement with the experimental measurement\cite{LHCb 2013}

$$\Gamma(B_{s2}^*(5840)\rightarrow B^{*+}K^-)/\Gamma(B_{s2}^*(5840)\rightarrow B^{+}K^-)=0.093\pm0.013\pm 0.012.$$

\subsubsection{Comparison to Other Work}

A summary of observables related to the recently observed  $B$ and $B_s$ mesons and  our predictions is
presented in Table \ref{experimental}. This table also shows related predictions from Ref. \cite{godfrey 2016}. This work suggests that the $B_J(5970)$ could be $2{}^3S_1$ state with a  predicted width of $107.8$ MeV. We have rather assigned this meson to the $1{}^3D_1$ state


%

\begin{table}[H]
\centering
\tabcolsep=1pt\fontsize{10}{10}\selectfont
\renewcommand{\arraystretch}{0.8}
\caption{Comparison of theoretical and experimental decay widths of observed $B$ and $B_s$ states. Experimental decay widths are taken from PDG \cite{PDG2016}, ratios of $B$ and $B_s$ states are taken from LHCb \cite{LHCb 2016} and \cite{LHCb 2013} respectively. Decay modes in brackets are "possibly seen".}
\begin{tabular}{c c c c c}
\hline\hline
State & Observed decays & Measured  & Our predicted & Ref. \cite{godfrey 2016} \\
      &                 &  $\Gamma$ (MeV)     &   $\Gamma$ (MeV) &  $\Gamma$ (MeV)\\
\hline
$B_1(5721)^+$        &  $B^{*0}\pi^+$    &$31\pm 6$       & 16.4   & 7.27\\
$B_1(5721)^0$        &  $B^{*+}\pi^-$    &$27.5\pm 3.4$   &        &     \\
\hline
$B^*_2(5747)^+$      &  $B^{0}\pi^+$, $B^{*0}\pi^+$   &$20\pm5$    & 20.3 & 11.71\\
$B^*_2(5747)^0$      &  $B^{+}\pi^-$, $B^{*+}\pi^-$   &$24.2\pm1.7$&      &      \\
                     & $\Gamma(B_2^{*+}\rightarrow B^{*0}\pi^+)/\Gamma(B_2^{*+}\rightarrow B^{0}\pi^+)$ & $1.0\pm0.5\pm0.8$    & 1.002 & 0.809 \\
                     & $\Gamma(B_2^{*0}\rightarrow B^{*+}\pi^-)/\Gamma(B_2^{*0}\rightarrow B^{+}\pi^-)$ & $0.71\pm0.14\pm0.3$    &  & \\
\hline
$B_J(5840)^+$        &  $[B^{0}\pi^+]$, $B^{*0}\pi^+$   &$224\pm80$   & 16  & 95 \\
$B_J(5840)^0$        &  $[B^{+}\pi^-]$, $B^{*+}\pi^-$   &$127\pm40$ \\
\hline
$B_J(5970)^+$        &  $[B^{0}\pi^+]$, $B^{*0}\pi^+$   &$62\pm20$   & 66.7  & 107.8 \\
$B_J(5970)^0$        &  $[B^{+}\pi^-]$, $B^{*+}\pi^-$   &$81\pm12$ \\
\hline
$B_{s1}(5830)^0$     &  $B^{*+}K^-$   &$0.5\pm0.4$    &kinematically excluded & no strong decay\\
\hline
$B^*_{s2}(5840)^0$   &  $B^{+}K^-$    &$1.47\pm0.33$  &1.9   & 0.777\\
                     & $\Gamma(B_{s2}^{*0}\rightarrow B^{*+}K^-)/\Gamma(B_{s2}^{*0}\rightarrow B^{+}K^-)$ & $0.093\pm0.013\pm0.012$    &0.084  &0.012 \\
\hline\hline
\end{tabular}
\label{experimental}
\end{table}

\subsubsection{$B_s$ meson decay properties}

\begin{table}[H]
\centering
\renewcommand{\arraystretch}{0.8}
\caption{Partial widths and branching ratios for strong, E1 and M1 decays of the 1S and 1P states of $B_s$ mesons (format as in Table \ref{d-1}).}
\begin{tabular}{c c c c c c}
\hline\hline
State &  Mode & Photon energy & Amplitude    & $\Gamma_{thy}$ & B.R \\
      &       & $(MeV)$       &              & $MeV$                        &($\%$) \\
\hline
$B^*_s$             & $B_s\gamma$&  48.41& $\langle 1^1S_0|j_0(kr\frac{m_{b,s}}{m_s+m_b})|1^3S_1\rangle=0.99469,0.9976$ & 0.0002& 100\\
\hline
$B_s(1 ^3P_0)$      & $B^*_s\gamma$& 343.4 & $\langle 1^3S_1|r|1^3P_0\rangle=3.0211$ & 0.133 & 0.0979\\
                    & $K B$   &         & $^1S_0=-0.764$ & 135.66& $\sim$ 100\\
                    & total   &         &                &  135.8     & 100\\
\hline
$B_s(1 ^3P_2)$      & $B^*_s\gamma$&  409.13 & $\langle 1^3S_1|r|1^3P_2\rangle=3.0211$ & 0.225 & 11.84\\
                    & $K B$   &         & $^1D_2=-0.044$ & 1.55  & 81.57\\
                    & $K B^*$ &         & $^3D_2=+0.018$ &0.13   & 6.84\\
                    &total    &         &                &1.9 & 100          \\
\hline\hline
\end{tabular}
\label{d-10}
\end{table}

\begin{table}[H]
\centering
\renewcommand{\arraystretch}{0.8}
\caption{Partial widths and branching ratios for strong, E1 and M1 decays of the 1D state of $B_s$ mesons (format as in Table \ref{d-1}).}
\begin{tabular}{c c c c c c}
\hline\hline
State &  Mode & Photon energy & Amplitude    & $\Gamma_{thy}$ & B.R \\
      &       & $(MeV)$       &              & $MeV$                        &($\%$) \\
\hline
$B_s(1 ^3D_1)$    & $B_s(1^3P_0)\gamma$&  280.33 & $\langle 1^3P_0|r|1^3D_1\rangle=4.99789$ & 0.126 & 0.1041\\
                  & $B_s(1P_1)\gamma$&  249.81 & $\langle 1^3P_1|r|1^3D_1\rangle=4.99789$ & 0.031 & 0.0256\\
                  & $B_s(1P'_1)\gamma$&  223.82 & $\langle 1^3P_1|r|1^3D_1\rangle=4.99789$ & 0.0264 & 0.0218\\
                  & $B_s(1^3P_2)\gamma$&  212.98 & $\langle 1^3P_2|r|1^3D_1\rangle=4.99789$ & 0.003 & 0.0025 \\
                  &$K B$     &       & $^1P_1=-0.174$ &69.8 & 57.65 \\
                  &$K B^*$   &       & $^3P_1=+0.135$ &36.27& 29.96\\
                  &$\eta B_s$&       & $^1P_1=-0.085$ &10.70& 8.8\\
                  &$\eta B_s^*$&     & $^3P_1=+0.060$ &4.11 & 3.39\\
                  &total     &       &                &121.06& 100   \\
\hline
$B_s(1 ^3D_3)$    & $B_s(1^3P_2)\gamma$&  218.87 & $\langle 1^3P_2|r|1^3D_3\rangle=4.99789$ & 0.109 & 0.5 \\
                  &$K B$     &       & $^1F_3=+0.067$ &10.64 & 48.8 \\
                  &$K B^*$   &       & $^3F_3=-0.072$ &10.45 & 47.9\\
                  &$\eta B_s$&       & $^1F_3=+0.017$ &0.44  & 2.01\\
                  &$\eta B_s^*$&     & $^3F_1=-0.013$ &0.19  & 0.87\\
                  &total     &       &                &21.8 & 100  \\
\hline
$B_s(1 D_2)$      & $B_s(1^3P_2)\gamma$&  217.22 & $\langle 1^3P_2|r|1^3D_2\rangle=4.99789$ & 0.027 & 0.1452 \\
                  &$K B^*$   &       & $^3P_2=+0.147c_{1}-0.180s_{1}$, & 17.89 & 96.18 \\
                  &          &       & $^3F_2=+0.073c_{1}+0.060s_{1}$  &        &   \\
                  &$\eta B_s^*$&     & $^3P_2=+0.067c_{1}-0.106s_{1}$, &0.66   & 3.54 \\
                  &          &       & $^3F_2=+0.013c_{1}+0.013s_{1}$  &        & \\
                  &total     &       &                                 &18.6  & 100 \\
\hline
$B_s(1 D_2^{'})$  & $B_s(1^3P_2)\gamma$&  218.13 & $\langle 1^3P_2|r|1^3D_2\rangle=4.99789$ & 0.0001 & 0.0001\\
                  &$K B^*$   &       & $^3D_2=-0.180c_{1}-0.146s_{1}$, &108.4   & 85.69 \\
                  &          &       & $^3F_2=+0.060c_{1}-0.073s_{1}$  &        & \\
                  &$\eta B_s^*$&     & $^3D_2=-0.106c_{1}-0.067s_{1}$, &18.13    & 14.33\\
                  &          &       & $^3F_2=+0.014c_{1}-0.013s_{1}$  &        & \\
                  &total     &       &                                 &126.5   & 100       \\
\hline\hline
\end{tabular}
\label{d-10}
\end{table}

\begin{table}[H]
\centering
\renewcommand{\arraystretch}{0.8}
\caption{Partial widths and branching ratios for strong, E1 and M1 decays of the 1F state of $B_s$ mesons (format as in Table \ref{d-1}).}
\begin{tabular}{c c c c c c}
\hline\hline
State &  Mode & Photon energy & Amplitude    & $\Gamma_{thy}$ & B.R \\
      &       & $(MeV)$       &              & $MeV$                        &($\%$) \\
\hline
$B_s(1 ^3F_2)$  & $B_s(1^3D_1)\gamma$&  212.64 & $\langle 1^3D_1|r|1^3F_2\rangle=6.54008$ & 0.155 & 0.2712\\
                & $B_s(1D_2)\gamma$&  208.39 & $\langle 1^3D_2|r|1^3F_2\rangle=6.54008$ & 0.027 & 0.0472\\
                & $B_s(1D'_2)\gamma$&  207.48 & $\langle 1^3D_2|r|1^3F_2\rangle=6.54008$ & 0.00013 & 0.0002\\
                & $B_s(1^3D_3)\gamma$&  206.74 & $\langle 1^3D_3|r|1^3F_2\rangle=6.54008$ & 0.0008 & 0.0014\\
                &$K B$    &        & $^1D_2=-0.067$ &17.74 & 31.04 \\
                &$K B^*$  &        & $^3D_2=-0.061$ &13.47 & 23.57 \\
                &$K^* B$  &        & $^3D_2=-0.081$ &13.75 & 24.05 \\
                &$K^* B^*$&        & $^1D_2=+0.044$, $^5D_2=-0.033$, $^5G_2=-0.014$ &4.73 & 8.27 \\
                &$\eta B_s$&       & $^1D_2=-0.037$ &4.22  & 7.38 \\
                &$\eta B_s^*$&     & $^3D_2=-0.033$ &3.06  & 5.35 \\
                &total    &        &                &57.15 & 100  \\
\hline
$B_s(1 ^3F_4)$  & $B_s(1^3D_3)\gamma$&  200.54 & $\langle 1^3D_3|r|1^3F_4\rangle=6.54008$ & 0.155 & 0.3444\\
                &$K B$    &        & $^1G_4=+0.051$ &10.34 & 22.97\\
                &$K B^*$  &        & $^3G_4=-0.060$ &12.79 & 28.4 \\
                &$K^* B$  &        & $^3G_4=-0.016$ &0.51  & 1.1333 \\
                &$K^* B^*$&        & $^5D_4=-0.117$, $^1G_4=+0.003$, $^5G_4=-0.006$ &18.73 & 41.6 \\
                &$\eta B_s$&       & $^1G_4=+0.021$ &1.30  & 2.89 \\
                &$\eta B_s^*$&     & $^3G_4=-0.021$ &1.19  & 2.64 \\
                &total    &        &                &45 & 100  \\
\hline\hline
\end{tabular}
\label{d-11}
\end{table}

\begin{table}[H]
\centering
\renewcommand{\arraystretch}{0.8}
\caption{Partial widths and branching ratios for strong, E1 and M1 decays of the 2S state of $B_s$ mesons (format as in Table \ref{d-1}).}
\begin{tabular}{c c c c c c}
\hline\hline
State &  Mode & Photon energy & Amplitude    & $\Gamma_{thy}$ & B.R \\
      &       & $(MeV)$       &              & $MeV$                        &($\%$) \\
\hline
$B_s(2 ^1S_0)$  &$B_s^*\gamma$ &  491.06   &$\langle 1^3S_1|j_0(kr\frac{m_{b,s}}{m_s+m_b})|2^1S_0\rangle=0.16581,-0.053989$ & 0.014 & 0.0358\\
                &$B_s(1P_1)\gamma$ & 125.69     &$\langle 1^1P_1|r|2^1S_0\rangle=-4.11746$ & 0.019 &0.0486 \\
                &$B_s(1P'_1)\gamma$ & 99.1    &$\langle 1^1P_1|r|2^1S_0\rangle=-4.11746$ & 0.008 & 0.0205 \\
                &$K B^*$   &       & $^3P_0=-0.267$  &39.03 & $\sim$ 100\\
                &total     &       &                 &39.1      & 100 \\
\hline
$B_s(2 ^3S_1)$  &$B_s\gamma$ & 553.69    &$\langle 1^1S_0|j_0(kr\frac{m_{b,s}}{m_s+m_b})|2^3S_1\rangle=0.24969,0.056964$ & 0.0167 & 0.0407\\
                &$B_s(2^1S_0)\gamma$ & 20.24    &$\langle 2^1S_0|j_0(kr\frac{m_{b,s}}{m_s+m_b})|2^3S_1\rangle=0.9946,0.99686$ & 0.00001 & $\sim$ 0\\
                &$B_s(1^3P_2)\gamma$ &  107.99   &$\langle 1^3P_2|r|2^3S_1\rangle=-4.00934$ & 0.0119 & 0.029\\
                &$B_s(1P_1)\gamma$ & 145.5    &$\langle 1^3P_1|r|2^3S_1\rangle=-4.00934$ & 0.008 & 0.0195\\
                &$B_s(1P'_1)\gamma$ & 119.03    &$\langle 1^3P_1|r|2^3S_1\rangle=-4.00934$ & 0.005 & 0.0122\\
                &$B_s(1^3P_0)\gamma$ & 176.57    &$\langle 1^3P_0|r|2^3S_1\rangle=-4.00934$ & 0.01 & 0.0244\\
                &$K B$     &       & $^1P_1=+0.118$  &10.74 & 26.19\\
                &$K B^*$   &       & $^3P_1=+0.203$  &25.49 & 62.17\\
                &$\eta B_s$&       & $^1P_1=+0.052$  &2.72  & 6.6\\
                &$\eta B_s^*$&     & $^3P_1=+0.059$  &2.07  & 5.04\\
                &total     &       &                 &41 & 100 \\
\hline\hline
\end{tabular}
\label{d-12}
\end{table}

\begin{table}[H]
\centering
\renewcommand{\arraystretch}{0.8}
\caption{Partial widths and branching ratios for strong, E1 and M1 decays of the 2P state of $B_s$ mesons (format as in Table \ref{d-1}).}
\begin{tabular}{c c c c c c}
\hline\hline
State &  Mode & Photon energy & Amplitude    & $\Gamma_{thy}$ & B.R \\
      &       & $(MeV)$       &              & $MeV$                        &($\%$) \\
\hline
$B_s(2 ^3P_0)$      &$B^*_s\gamma$ & 699.23    &$\langle 1^3S_1|r|2^3P_0\rangle=0.528051$ & 0.031 & 0.0449\\
                    &$B_s(2^3S_1)\gamma$ &  207.03   &$\langle 2^3S_1|r|2^3P_0\rangle=4.82565$ & 0.072 & 0.1042\\
                    &$B_s(1^3D_1)\gamma$ & 101.9    &$\langle 1^3D_1|r|2^3P_0\rangle=-4.24042$ & 0.014 & 0.0203\\
                    & $K B$     &      & $^1S_0=-0.144$ &63.90 & 92.47 \\
                    & $\eta B_s$&      & $^1S_0=-0.047$ &5.09  & 7.36\\
                    &total      &      &                &69.1 & 100\\
\hline
$B_s(2 P_1)$        &$B_s\gamma$ &  768.11   &$\langle 1^1S_0|r|2^1P_1\rangle=0.60031$ & 0.028 & 0.5469\\
                    &$B^*_s\gamma$ & 725.75    &$\langle 1^3S_1|r|2^3P_1\rangle=0.52805$ & 0.016 & 0.3125\\
                    &$B_s(2^1S_0)\gamma$ &  255.39   &$\langle 2^1S_0|r|2^1P_1\rangle=4.5595$ & 0.065 & 1.2695\\
                    &$B_s(2^3S_1)\gamma$ & 235.94    &$\langle 2^3S_1|r|2^3P_1\rangle=4.8256$ & 0.048 & 0.9375\\
                    & $K B^*$   &      & $^3S_1=-0.082c+0.115s$ &2.71 & 52.9 \\
                    &           &      & $^3D_1=-0.024c-0.018s$ &      & \\
                    & $\eta B_s^*$&    & $^3S_1=-0.027c+0.049s$ &2.25  & 43.9 \\
                    &           &      & $^3D_1=-0.024c-0.021s$ &      & \\
                    &total      &      &                        &5.12 & 100\\
\hline
$B_s(2 P_1^{'})$    &$B_s\gamma$ &  771.09   &$\langle 1^1S_0|r|2^1P_1\rangle=0.60031$ & 0.024 & 0.0359\\
                    &$B^*_s\gamma$ & 729.11    &$\langle 1^3S_1|r|2^3P_1\rangle=0.52805$ & 0.019 &0.0284 \\
                    &$B_s(2^1S_0)\gamma$ &  258.65   &$\langle 2^1S_0|r|2^1P_1\rangle=4.5595$ & 0.057 & 0.0852\\
                    &$B_s(2^3S_1)\gamma$ & 239.21    &$\langle 2^3S_1|r|2^3P_1\rangle=4.8256$ & 0.06 & 0.0897\\
                    & $K B^*$   &      & $^3S_1=+0.117c+0.083s$ &60.28 & 90.1 \\
                    &           &      & $^3D_1=-0.018c+0.023s$ &     &\\
                    & $\eta B_s^*$&    & $^3S_1=+0.049c+0.026s$ &6.46 & 9.65 \\
                    &           &      & $^3D_1=-0.021c+0.024s$ &     & \\
                    &total      &      &                        & 66.9& 100\\
\hline
$B_s(2 ^3P_2)$      &$B^*_s\gamma$ & 742.3    &$\langle 1^3S_1|r|2^3P_2\rangle=0.52805$ & 0.036 & 0.309\\
                    &$B_s(2^3S_1)\gamma$ &  253.97   &$\langle 2^3S_1|r|2^3P_2\rangle=4.8256$ & 0.132 & 1.133 \\
                    &$B_s(1^3D_1)\gamma$ & 149.67    &$\langle 1^3D_1|r|2^3P_2\rangle=-4.2404$ & 0.0004 & 0.0034\\
                    &$B_s(1^3D_3)\gamma$ &  143.71   &$\langle 1^3D_3|r|2^3P_2\rangle=-4.2404$ & 0.032 & 0.2747\\
                    & $K B$     &      & $^1D_2=-0.041$ &5.91 & 50.72\\
                    & $K B^*$   &      & $^3D_2=+0.030$ &2.65 & 22.7\\
                    & $K^* B$   &      & $^3D_2=+0.037$ &1.62 & 13.9\\
                    & $\eta B_s$&      & $^1D_2=-0.011$ &0.29 & 2.48\\
                    & $\eta B_s^*$&    & $^3D_2=+0.021$ &0.98 & 8.4\\
                    &total      &      &                &11.65& 100          \\
\hline\hline
\end{tabular}
\label{d-13}
\end{table}

\begin{table}[H]
\centering
\renewcommand{\arraystretch}{0.8}
\caption{Partial widths and branching ratios for strong, E1 and M1 decays of the 2D state of $B_s$ mesons (format as in Table \ref{d-1}).}
\begin{tabular}{c c c c c c}
\hline\hline
State &  Mode & Photon energy & Amplitude    & $\Gamma_{thy}$ & B.R \\
      &       & $(MeV)$       &              & $MeV$                        &($\%$) \\
\hline
$B_s(2 ^3D_1)$    &$B_s(1^3P_0)\gamma$ & 590.32    &$\langle 1^3P_0|r|2^3D_1\rangle=0.44889$ & 0.009 & 0.0157\\
                  &$B_s(1^3P_2)\gamma$ & 526.48    &$\langle 1^3P_2|r|2^3D_1\rangle=0.44889$ &0.0003  & 0.0005\\
                  &$K B$      &       & $^1P_1=-0.058$ &16.60 & 28.86\\
                  &$K B^*$    &       & $^3P_1=+0.045$ &9.29  & 16.15\\
                  &$K^* B$    &       & $^3P_1=+0.057$ &11.02 & 19.16\\
                  &$K^* B^*$  &       & $^1P_1=+0.038$,$^5P_1=-0.041$,$^5F_1=-0.048$ &15.81 & 27.49\\
                  &$\eta B_s$ &       & $^1P_1=-0.028$ &3.11  & 5.4\\
                  &$\eta B_s^*$&      & $^3P_1=+0.021$ &1.67  & 2.9\\
                  &total      &       &                &57.5  & 100 \\
\hline
$B_s(2 ^3D_3)$    &$B_s(1^3P_2)\gamma$ & 534.19    &$\langle 1^3P_2|r|2^3D_3\rangle=0.44889$ & 0.012 & 0.0147\\
                  &$K B$      &       & $^1F_3=+0.066$ &21.65 & 26.6 \\
                  &$K B^*$    &       & $^3F_3=-0.069$ &21.84 & 26.83\\
                  &$K^* B$    &       & $^3F_3=-0.007$ &0.15  & 0.18\\
                  &$K^* B^*$  &       & $^5P_3=-0.108$,$^1F_3=+0.011$,$^5F_3=-0.024$ &36.79 & 45.2\\
                  &$\eta B_s$ &       & $^1F_3=+0.012$ &0.63  & 0.77\\
                  &$\eta B_s^*$&      & $^3F_1=-0.009$ &0.31  & 0.38\\
                  &total      &       &                &81.38 & 100  \\
\hline
$B_s(2 D_2)$      &$B_s(1^3P_2)\gamma$ & 531.58    &$\langle 1^3P_2|r|2^3D_2\rangle=0.44889$ & 0.003 & 0.0079\\
                  &$B_s(2^3P_2)\gamma$ & 184.6    &$\langle 2^3P_2|r|2^3D_2\rangle=6.5948$ & 0.029 & 0.0759\\
                  &$K B^*$    &       & $^3P_2=+0.048c_{1}-0.059s_{1}$, &37.47 & 98.09 \\
                  &           &       & $^3F_2=+0.070c_{1}+0.057s_{1}$  &      &\\
                  &$\eta B_s^*$&      & $^3P_2=+0.023c_{1}-0.037s_{1}$, &0.73  & 1.9 \\
                  &           &       & $^3F_2=+0.009c_{1}+0.009s_{1}$  &      & \\
                  &total      &       &                                 &38.2 & 100  \\
\hline
$B_s(2 D_2^{'})$  &$B_s(1^3P_2)\gamma$ & 532.11    &$\langle 1^3P_2|r|2^3D_2\rangle=0.44889$ & 0.00001 & $\sim$0 \\
                  &$B_s(2^3P_2)\gamma$ &  185.16   &$\langle 2^3P_2|r|2^3D_2\rangle=6.5948$ & 0.0001 & 0.0003\\
                  &$K B^*$    &       & $^3D_2=-0.059c_{1}-0.048s_{1}$, &26.65 & 79.31 \\
                  &           &       & $^3F_2=+0.057c_{1}-0.070s_{1}$  &      &\\
                  &$\eta B_s^*$&      & $^3D_2=-0.037c_{1}-0.023s_{1}$, &6.98  & 20.77 \\
                  &           &       & $^3F_2=+0.010c_{1}-0.009s_{1}$  &      & \\
                  &total      &       &                                 &33.6 & 100         \\
\hline\hline
\end{tabular}
\label{d-14}
\end{table}

\begin{table}[H]
\centering
\renewcommand{\arraystretch}{0.8}
\caption{Our assignments for the observed $B$ and $B_s$ states are based on their masses, $J^P$ and decay widths. Experimental masses and $J^P$ for the states are taken from PDG \cite{PDG2016} and
theoretical masses are calculated by using a potential model discussed in Sec.(\ref{spectrum}). In the 4th column, there are two experimental masses, first for $B^0(d\overline{b})$ and second for $B^+(u\overline{b})$.}
\begin{tabular}{c c c c c}
\hline\hline
State & $J^P$ & Our assignment & Experimental mass & Theoretical mass \\
      &       &                & (MeV) & (MeV) \\
\hline
$B_1(5721)$      &  $1^+$    &$B(1P_1)$ & $5726.0\pm 1.3$      &5755.12\\
                 &           &          & $5725.9^{+2.5}_{-2.7}$\\
\hline
$B^*_2(5747)$    &  $2^+$    &$B(1^3P_2)$& $5739.5\pm 0.7$     &5768.95\\
                 &           &          & $5737.2\pm 0.7$\\
\hline
$B_J(5970)$      &   ...    &$B(1^3D_1)$ & $5971\pm 5$ & 6022.39 \\
                 &           &          & $5964\pm 5$\\
\hline
$B_{s1}(5830)$   &  $1^+$    &$B_s(1P_1)$ &$5828.63\pm 0.27$ & 5800.95 \\
\hline
$B^*_{s2}(5840)$ &  $2^+$    &$B_s(1^3P_2)$ & $5839.84\pm 0.18$ & 5821.53 \\
\hline\hline
\end{tabular}
\label{assignment}
\end{table}

\section{Conclusions}\label{conclusions}

We have computed the spectrum of $B$ and $B_s$ mesons up to $2F$ states with a nonrelativistic quark model that incorporates ``scalar" confinement and one gluon exchange spin-dependent interactions. The eigenfunctions were then used to obtain E1 and M1 radiative transitions using the nonrelativistic reduction of the transition amplitude in impulse approximation. Strong decay amplitudes have also been obtained using the ${}^3P_0$ pair creation model and fitted SHO wavefunctions.

A comparison to four recently found $B$ and $B_s$ mesons with known $J^P$ reveals good agreement with their natural candidate states in terms of mass and measured strong decay widths. The $B_J(5840)$ and $B_J(5970)$ are tougher to identify because their spin-parity has not been measured. However, we find that the lighter signal matches well with expectations for a  $2^1S_0$ state, while the $B_J(5970)$ is  most naturally identified with the $1^3D_1$ quark model state.

At present, excited states of the $B$ and $B_s$ mesons beyond P-wave have not been identified. We nevertheless expect that the LHCb collaboration will play a very important role in the study of excited states of the $B$ and $B_s$ meson families. The spectrum, radiative transitions, and  strong decay widths presented here should prove helpful in the search for, and interpretation of, these states.

\section{Acknowledgement}
B. Masud and F. Akram acknowledge the financial support of Punjab University.

\begin{appendix}
\section{SHO $\beta$ values and masses for light mesons} \label{light mesons}
\indent The light meson masses used to determine phase space and final state momenta are \cite{PDG2014}: $\pi=0.1372$, $\eta=0.5478$, $\rho=0.7752$, $\omega=0.7826$, $K=0.495$, $K^*=0.894$, $\phi=1.019$. The SHO $\beta$ values used in this work are listed in Table \ref{beta table 1} and \ref{beta table 2} which are obtained by using the technique described in Sec. \ref{sect:decay}.
\begin{table}[H]
\centering
\renewcommand{\arraystretch}{0.7}
\caption{Fitted $\beta$ values (GeV) for $ub$ and $sb$.}
\begin{tabular}{c c c c c c }
\hline\hline
$n\;^{2S+1}L_J$ \hspace{0.8cm} & \hspace{0.8cm}$ub$\hspace{0.8cm} & \hspace{0.8cm} $sb$ \hspace{0.8cm}  \\
\hline
$1^1S_0$ & 0.405 & 0.429  \\
$1^3S_1$ & 0.372 & 0.401  \\
$1^1P_1$ & 0.295 & 0.318  \\
$1^3P_J$ & 0.292 & 0.316 \\
$1^1D_2$ & 0.264 & 0.286  \\
$1^3D_J$ & 0.264 & 0.286 \\
$1^1F_3$ & 0.248 & 0.269  \\
$1^3F_J$ & 0.248 & 0.269  \\
$2^1S_0$ & 0.323 & 0.346  \\
$2^3S_1$ & 0.309 & 0.334  \\
$2^1P_1$ & 0.271 & 0.293  \\
$2^3P_J$ & 0.270 & 0.292  \\
$2^1D_2$ & 0.251 & 0.272  \\
$2^3D_J$ & 0.251 & 0.272  \\
$3^1S_0$ & 0.280 & 0.301  \\
$3^3S_1$ & 0.273 & 0.295  \\
$3^1P_1$ & 0.252 & 0.272  \\
$3^3P_J$ & 0.251 & 0.271  \\
$3^1D_2$ & 0.239 & 0.258  \\
$3^3D_J$ & 0.238 & 0.258  \\
$4^1S_0$ & 0.256 & 0.276  \\
$4^3S_1$ & 0.252 & 0.272  \\
\hline\hline
\end{tabular}
\label{beta table 1}
\end{table}

\begin{table}[H]
\centering
\renewcommand{\arraystretch}{0.8}
\caption{Fitted $\beta$ values (GeV) for light mesons.}
\begin{tabular}{c c c c}
\hline\hline
$n\;^{2S+1}L_J$ \hspace{0.8cm}& \hspace{0.8cm}$uu$ \hspace{0.8cm} & \hspace{0.4cm}$us$ \hspace{0.4cm}& \hspace{0.6cm} $ss$ \hspace{0.8cm} \\
\hline
$1^1S_0$ & 0.489 & 0.506 & 0.337\\
$1^3S_1$ & 0.200 & 0.324 & 0.337\\
\hline\hline
\end{tabular}
\label{beta table 2}
\end{table}

\end{appendix}

\end{document}